\documentclass[letterpaper,twocolumn,showpacs,
preprintnumbers,amsmath,amssymb,floatfix,nofootinbib]{revtex4}

\usepackage{graphicx}
\usepackage{bm}
\usepackage{url,multirow,xcolor}

\begin{document}

\title{ Overlap of QRPA states based on ground states of different nuclei

--mathematical properties and test calculations-- }

\author{J.\ Terasaki}
\affiliation{Division of Physics and Center for Computational Sciences, University of Tsukuba, Tsukuba 305-8577, Japan}

\begin{abstract} 
The overlap of the excited states in quasiparticle random-phase approximation (QRPA) is calculated 
in order to simulate the overlap of the intermediate nuclear states of the double-beta decay. 
Our basic idea is to use the like-particle QRPA with the aid of the closure approximation and calculate the overlap  as rigorously as possible by making use of the explicit equation of the QRPA ground state. 
The formulation is shown in detail, and the mathematical properties of the overlap matrix are investigated. 
Two test calculations are performed for relatively light nuclei with the Skyrme and volume delta-pairing energy functionals. 
The validity of the truncations used in the calculation is examined and confirmed. 
\end{abstract}

%
\maketitle
\section{\label{sec:introduction}Introduction}
One of features of neutrino physics is its interdisciplinarity. Neutrino physics is important for a better understanding of particle physics 
in terms of lepton-number violation, Majorana nature of neutrino, and neutrino mass, e.g.~Refs.~\cite{Moh10,Bil87,Doi85}, (there are many textbooks elucidating 
neutrino; see for example Ref.~\cite{Boe92}), which are aspects of particle physics beyond the scope of the standard model. 
Neutrino physics is also very interesting from the viewpoint of nuclear physics \cite{Fae12,Ver02,Fae98,Suh98,Sim98,Pan94,Tom91,Gro90,Hax84}. 
One of the few methods used to determine the neutrino mass requires 
accurate calculations of the nuclear matrix elements in the neutrino-less double-beta (0$\nu\beta\beta$) decay along with its experimental half life,  e.g.~Refs.~\cite{Doi85,Hax84}. 
For details on other methods to determine the neutrino mass, see e.g.~Ref.~\cite{Wil10} (shape of the $\beta$ decay spectra and other particle-physical methods) 
and Ref.~\cite{Les06} (cosmological method).
The primary task is to determine the neutrino mass accurately, on the other hand, this is a very good and challenging opportunity 
for theoretical nuclear physics to test if the techniques developed so far in this field are useful for 
solving the problem of other field. This is particularly because most of the nuclei providing the ground of the $0\nu\beta\beta$ decay are  
heavy nuclei, and therefore many-body correlations have to be taken into account along with large wave-function space. 

In this study, we take the first step\footnote{Parts of the formulation and the test calculations in this paper are reported in Ref.~\cite{Ter12}.} in the attempt to calculate the nuclear matrix elements of the $0\nu\beta\beta$ decay by making use of a method different from the traditional ones; First, we use the like-particle quasiparticle random-phase approximation (QRPA) \cite{Rin80} (formulation of the QRPA) 
\cite{Vog10} (application to the $0\nu\beta\beta$ decay suggested) for axially-deformed nuclei, which can be applied after the closure approximation is used. This approximation has been proven to be good in the $0\nu\beta\beta$ decay by the analytical argument \cite{Tom91} 
and several realistic calculations \cite{Fae91,Suh91,Suh90,Pan90,Hax84}.

In the application of the QRPA to the $0\nu\beta\beta$ decay, two QRPA-state spaces are obtained via calculations based on the initial and final states of the decay, and 
the product of the two projection operators to the QRPA-state spaces is inserted to the middle of the two-body $0\nu\beta\beta$ transition operator.
Secondly, in our approach, the overlap of the intermediate states obtained by the two QRPA calculations is calculated more accurately than ever. 
A simple approximation and a few variants \cite{Sim04,Kam91,Sim98,Civ86,Gro85} have been used for calculating the overlap. 
The importance of the overlap of the intermediate
states is pointed out in Ref.~\cite{Sim04} in terms of deformation. It is reasonable that the overlap is sensitive to the difference in the deformation of the initial and final states; this raises a question if the differences in other properties affect the overlap. We can address this question comprehensively by treating the ground-state wave function of the QRPA {\it explicitly}, and here we demonstrate the feasibility of that treatment and investigate mathematical properties of the overlap. 
The equation of the QRPA ground state has been known for decades, e.g.~Ref.~\cite{Bal69}; 
however, to the best of our knowledge, our study is the first that carries out the involved numerical calculation rigorously. 
Few researchers have attempted to calculate explicitly the QRPA ground state in subjects other than the study of the $0\nu\beta\beta$ decay  \cite{Bal69,Dap66,Bro63}, and they have mostly used crude approximations. 
The probable reason for the rarity in attempting this calculation is that it is possible to obtain 
the transition strength of the QRPA without treating the explicit ground-state wave function \cite{Rin80}. In this light, neutrino physics provides  further motivation to develop techniques of nuclear theory. 

The third feature of our approach is the use of the Skyrme energy density functional \cite{Sky56a,Sky56b}. 
It is of interest from the viewpoint of nuclear theory 
to investigate how a phenomenological approach developed so as to reproduce as many experimental data as possible including the masses and the root-mean-square radii of the ground states can be successful in describing other nuclear properties. The Skyrme energy density functional has been used
for providing the Hartree-Fock field to the calculation of the nuclear matrix elements \cite{Mor09}. We use the Skyrme-plus-pairing energy density functional to solve the Hartree-Fock-Bogoliubov (HFB) \cite{Rin80} and the subsequent QRPA equations self-consistently. The self-consistency assists in strengthening the reliability of the calculation.

The standard method in the category of the QRPA for calculating the nuclear matrix elements is the proton-neutron (pn) QRPA \cite{Hal67}. It has been argued that the Pauli correction terms are necessary to include in the calculation of the intermediate states. The renormalized pn-QRPA  \cite{Har64,Row68,Cat94} has been used for including the Pauli correction terms, and later the Ikeda sum rule \cite{Ike64} was satisfied upon using the fully renormalized pn-QRPA \cite{Pac03,Rod02}. 
The self-consistent HFB and pn-QRPA calculations have been performed in Ref.~\cite{Bob01}. 
The importance of the particle-particle interaction has been pointed out in Ref.~\cite{Eng88}, and subsequently   
the proton-neutron pairing correlations have also been included in Ref.~\cite{Sch96}. 
Further, a pn-QRPA calculation using the unitary correlation operator method has been performed for taking into account the short-range correlations \cite{Kor07}. 
As previously mentioned, another improvement as regards the pn-QRPA is its extension to deformed states \cite{Fan11,Rad93}. 
Thus, the pn-QRPA has been improved up to a very advanced level in the past few decades. 
Nevertheless, as is well known, the problem of the systematic difference in the nuclear matrix elements between different approaches has not thus  far been resolved \cite{Fae12,MED11}; in particular, there is a difference of a factor of two between the pn-QRPA and the shell-model approach. As for approaches other than the pn-QRPA, these can be found in  
e.g, Refs.~\cite{Suh97,Koo97,Cau96,Nak96,Zha90} (the shell-model), 
\cite{Rat10} (the projected HFB), \cite{Bar09} (the microscopic IBM), and \cite{Rod10} (the energy-density functional-plus-generator-coordinate method). 

One of the advantages of our method is that 
the feasibility of the like-particle QRPA calculation is fairly high for any nuclei except for the transitional ones between the spherical and deformed regions. To the best of our knowledge, the collapse of the like-particle QRPA due to the pairing fluctuation does not occur \cite{Ter06} as long as the strength of the pairing energy functional is determined so as to reproduce the pairing gaps obtained from experimental odd-even mass differences \cite{Boh69}.
Another advantage is that the calculation of even-even nuclei is free from a problem that the last odd particle may not be approximated very well by the HFB calculation, if the coupling of the last particle to the nucleus is weak. 
The drawback of our approach is that since the closure approximation is not good for the $2\nu\beta\beta$ decay, e.g.~\cite{Hax84}, the reliability of our method is difficult to  prove by itself. Perhaps it is necessary to rely on other methods to obtain a reference value of the representative energy of the intermediate states which makes the closure approximation exact.
If the effects of the higher-order many-body correlations beyond the QRPA are minor, 
and a sufficiently large space of the intermediate states is used, then 
the question is whether 
the pn-QRPA and the like-particle QRPA provide similar nuclear matrix elements. 
The answer is not trivial, because the many-body correlations treated in the two QRPA methods are different. Thus, it is worthy to compare the two methods  numerically.

This paper is organized as follows: Section \ref{sec:formulation} presents our basic scheme to calculate the nuclear matrix elements and the detailed formulation for calculating the overlap matrix elements of the intermediate states. 
The analytical properties of the overlap matrix are discussed in Sec.~\ref{sec:analytical} for simplified cases. 
Section \ref{sec:calculation} provides technical information regarding the numerical calculations for this paper.
The calculated overlap matrix elements are shown in Sec.~\ref{sec:result}, and the truncation approximations are examined in detail.
Section \ref{sec:summary} summarizes the study.

\section{\label{sec:formulation} Formulation}
The axial and parity symmetries of the nuclei are assumed throughout this paper.
The $z$-component of the angular momentum is denoted by $j^z_\alpha$ for nucleon state $\alpha$ and by $K_m$ for nuclear state $m$.
The terms $\pi_\alpha$ and $\pi_m$ are used to indicate the parity.
All of the matrix elements used in the numerical calculations of this paper are real, although it is not assumed in the formulation of this section.
Hereafter, we call the like-particle QRPA as simply the QRPA. 
For complete equations of the nuclear matrix elements, see e.g.~Refs.~\cite{Sim99,Doi85,Hax84}.

\subsection{\label{subsec:like-particle_QRPA} Application of QRPA to our method}
As mentioned in Sec.~\ref{sec:introduction}, we assume that the closure approximation has been applied to the nuclear matrix elements of 
the $0\nu\beta\beta$ decay.
One of the components of the nuclear matrix element arises from the double Gamow-Teller operator \cite{Doi85}, and this component is now written as
\begin{eqnarray}
M^{(0\nu)}_\textrm{GT} &=&
\langle F|\sum_{ij}h_+(r_{ij},\bar{E}_a)\bm{\sigma}(i)\cdot\bm{\sigma}(j)\tau_{+}(i)\tau_{+}(j) | I \rangle \nonumber \\
&=&\sum_{\alpha\beta}\sum_{\alpha^\prime\beta^\prime}
\langle \alpha \alpha^\prime | h_+(r_{12},\bar{E}_a)\bm{\sigma}(1)\cdot\bm{\sigma}(2) \nonumber \\
&&\times\tau_{+}(1)\tau_{+}(2) | \beta^\prime \beta\rangle
\langle F | c^\dagger_{\alpha} c^\dagger_{\alpha^\prime}c_\beta c_{\beta^\prime} | I \rangle ,
\label{eq:double_GT}
\end{eqnarray}
where $|F\rangle$ and $|I\rangle$ denote the final and initial nuclear states of the decay, and the ground states of the QRPA are used. 
$h_+(r_{ij},\bar{E}_a)$ is the neutrino potential \cite{Doi85} with $r_{ij}=|\bm{r}_i-\bm{r}_j|$, and 
$\bar{E}_a$ being the average energy of the intermediate nuclear states. 
The index $i$ $(j)$ indicates a nucleon,  
$\bm{\sigma}$ denotes the spin-Pauli matrix, and $\tau_+$ denotes the raising operator of the $z$-component of 
the isospin. An arbitrary single-particle basis $\{\alpha\}$ is introduced, 
and the creation and annihilation operators of the single-particle state are denoted by $c^\dagger_{\alpha}$ and $c_\alpha$, respectively.

We introduce the creation and annihilation operators $O^{I\dagger}_m$ and $O^I_m$, respectively, of the excited state $m$ of the QRPA based on the initial state,  
and those based on the final state are denoted by $O^{F\dagger}_m$ and $O^F_m$.
The same kind of index $m$ is used for specifying the QRPA states based on the initial or the final state.
The states $|F\rangle$ and $|I\rangle$ are defined in the QRPA by 
\begin{eqnarray}
&& O^I_m|I\rangle = 0, \\
&& O^F_m|F\rangle = 0.
\end{eqnarray}
Inserting the product of the two completeness equations of the relevant space
\begin{eqnarray}
1 &=& | I \rangle \langle I | + \sum_m O^{I\dagger}_m | I \rangle \langle I | O^I_m  \nonumber \\
&& +\sum_{m_1 m_2}O^{I\dagger}_{m_1} O^{I\dagger}_{m_2} | I \rangle \langle I | O^I_{m_2} O^I_{m_1} + \cdots ,\nonumber \\
1 &=& | F \rangle \langle F | + \sum_m O^{F\dagger}_m | F \rangle \langle F | O^F_m  \nonumber \\
&& +\sum_{m_1 m_2}O^{F\dagger}_{m_1} O^{F\dagger}_{m_2} | F \rangle \langle F | O^F_{m_2} O^F_{m_1} + \cdots ,
\label{eq:completeness}
\end{eqnarray}
to the middle of the product of the single-particle operators, we get
\begin{eqnarray}
M^{(0\nu)}_\textrm{GT} &=& \sum_{\alpha\beta}\sum_{\alpha^\prime\beta^\prime}
\langle \alpha \alpha^\prime| h_+(r_{12},\bar{E}_a)\bm{\sigma}(1)\cdot \bm{\sigma}(2) \nonumber\\
&&\times \tau_{+}(1) \tau_{+}(2) | \beta \beta^\prime \rangle 
\sum_{mm^\prime}\langle F | c^\dagger_{\alpha} c^\dagger_{\alpha^\prime} O^{F\dagger}_m | F \rangle \nonumber \\
&&\times\langle F | O^F_m O^{I\dagger}_{m^\prime} | I \rangle \langle I | O^I_{m^\prime} c_{\beta^\prime} c_{\beta} | I \rangle . \label{eq:double_GT_our_original}
\end{eqnarray}
We assume that the higher-order terms with respect to $O^{I(F)}_m$ or $O^{\dagger I(F)}_m$ do not have contribution to the two-particle transfer matrix element, for example
\begin{equation}
\langle I | O^I_{m_1} O^I_{m_2} c_{\beta^\prime} c_{\beta} | I \rangle = 0. \label{eq:forbidden_2particle_transfer}
\end{equation} 
(In the QRPA order, this equation holds exactly.)
The nuclear states $O^{F\dagger}_m|F\rangle$ and $O^{I\dagger}_{m^\prime}|I\rangle$ in Eq.~(\ref{eq:double_GT_our_original}) are 
the intermediate states mentioned, and   
the overlap of these two intermediate states is not equal to $\delta_{mm^\prime}$ in the QRPA.

The conditions for the product 
\begin{eqnarray} 
&&\langle F | c^\dagger_{\alpha} c^\dagger_{\alpha^\prime} O^{F\dagger}_{m} | F \rangle 
 \langle F | O^F_m O^{I\dagger}_{m^\prime} | I \rangle 
\langle I | O^I_{m^\prime} c_{\beta^\prime} c_{\beta} | I \rangle , \label{eq:factors}
\end{eqnarray}
in Eq.~(\ref{eq:double_GT_our_original}) to be finite are 
\begin{eqnarray}
&& j^z_\alpha + j^z_{\alpha^\prime} = j^z_\beta + j^z_{\beta^\prime}, \nonumber \\
&& \pi_\alpha \pi_{\alpha^\prime} = \pi_\beta \pi_{\beta^\prime} , \label{eq:condition_jz_pi}
\end{eqnarray}
for the single-particle states, and
\begin{eqnarray}
&& K_m = K_{m^\prime} = -j^z_\alpha  -j^z_{\alpha^\prime} , \nonumber \\
&& \pi_m = \pi_{m^\prime} =\pi_\alpha \pi_{\alpha^\prime} ,
\end{eqnarray}
for the intermediate states.
For an arbitrary pair of $\alpha$ and $\alpha^\prime$, there exist $\beta$ and $\beta^\prime$  satisfying condition (\ref{eq:condition_jz_pi}) and  the condition that the two-body matrix element of the double Gamow-Teller operator is finite.
Thus, the QRPA solutions are necessary for all $(K_m \pi_m)$ for which Eq.~(\ref{eq:double_GT_our_original}) is convergent; 
in other words, there is no selection rule for the intermediate states.

\subsection{\label{sec:overlap} Formulation of overlap of intermediate states}

In this subsection, we show the detailed equations for calculating the overlap matrix elements
\begin{equation}
_F\langle m | m^\prime\rangle_I 
\equiv \langle F | O^F_m O^{I\dagger}_{m^\prime} | I \rangle .
\end{equation}
Hereafter, we use the simplified notations $K=K_m$ and $\pi=\pi_m$.
We express $|I\rangle$ and $|F\rangle$ in the form \cite{Bal69},
\begin{eqnarray}
&& |I\rangle = \frac{1}{ {\cal N}_I} \prod_{K^\prime \pi^\prime} \exp\left[\hat{v}^{(K^\prime \pi^\prime)}_I\right]|i\rangle,\\
&& |F\rangle = \frac{1}{ {\cal N}_F }\prod_{K^\prime\pi^\prime} \exp\left[\hat{v}^{(K^\prime\pi^\prime)}_F\right]|f\rangle, 
\end{eqnarray}
where
$|i\rangle$ and $|f\rangle$ denote the HFB ground states of the nuclei described by $|I\rangle$ and $|F\rangle$, respectively, and 
$\hat{v}^{(K^\prime\pi^\prime)}_I$ and $\hat{v}^{(K^\prime\pi^\prime)}_F$ denote the generators of the QRPA ground states. 
The terms ${\cal N}_I$ and ${\cal N}_F$ indicate the normalization factors. 
We have $[O^\dagger_m,O_{m^\prime}]=0$ in the QRPA if $(K\pi)\neq (K_{m^\prime}\pi_{m^\prime})$, 
and hence, $\hat{v}^{(K^\prime\pi^\prime)}_I$'s and $\hat{v}^{(K^\prime\pi^\prime)}_F$'s with different values of $(K^\prime\pi^\prime)$ are determined separately by using
\begin{eqnarray}
&&O^I_{m^\prime} \exp \left[ \hat{v}^{(K_{m^\prime}\pi_{m^\prime})}_I \right] |i \rangle = 0, \\
&&O^F_{m^\prime} \exp \left[ \hat{v}^{(K_{m^\prime}\pi_{m^\prime})}_F \right] |f \rangle = 0.
\end{eqnarray}

General quasiparticle bases, which are not necessarily the diagonal representation of the HFB Hamiltonian, are introduced by using $|i\rangle$ and $|f\rangle$ as the vacuum state, that is,
\begin{eqnarray}
&& a^I_\mu|i\rangle = 0, \\
&& a^F_\mu|f\rangle = 0,
\end{eqnarray}
where $\mu=(q_\mu,\pi_\mu,j^z_\mu, i_\mu)$ denotes the label of a general quasiparticle state. The term $q_\mu$ indicates proton or neutron, and $i_\mu$ denotes a label specifying the general quasiparticle state in the subspace ($q_\mu,\pi_\mu,j^z_\mu$). 
The notation $-\mu$ is used for expressing 
$(q_\mu,\pi_\mu,-j^z_\mu,i_\mu)$.
The generators $\hat{v}^{(K^\prime\pi^\prime)}_I$ and $\hat{v}^{(K^\prime\pi^\prime)}_F$ are written as  
\begin{eqnarray}
&&\hat{v}^{(K^\prime\pi^\prime)}_I = \sum_{\mu\nu\mu^\prime\nu^\prime} C^{(K^\prime\pi^\prime)I}_{\mu\nu,\mu^\prime\nu^\prime}
 a^{I\dagger}_\mu a^{I\dagger}_\nu a^{I\dagger}_{\mu^\prime} a^{I\dagger}_{\nu^\prime}, \label{eq:vhatI}\\
&&\hat{v}^{(K^\prime\pi^\prime)}_F = \sum_{\mu\nu\mu^\prime\nu^\prime} C^{(K^\prime\pi^\prime)F}_{\mu\nu,\mu^\prime\nu^\prime} 
a^{F\dagger}_\mu a^{F\dagger}_\nu a^{F\dagger}_{\mu^\prime} a^{F\dagger}_{\nu^\prime}~. \label{eq:vhatF}
\end{eqnarray}
It is to be noted that $a^{I\dagger}_\mu a^{I\dagger}_\nu$ and $a^{I\dagger}_{\mu^\prime} a^{I\dagger}_{\nu^\prime}$ 
in Eq.~(\ref{eq:vhatI}) are the fermion images of bosons. 
In relation to this, we introduce the condition that $C^{(K^\prime\pi^\prime)I}_{\mu\nu,\mu^\prime\nu^\prime}$ does not vanish 
only if 
$j^z_\mu+j^z_\nu=K^\prime$, $j^z_{\mu^\prime}+j^z_{\nu^\prime}=-K^\prime$, and 
$\pi_\mu \pi_\nu= \pi_{\mu^\prime}\pi_{\nu^\prime}=\pi^\prime$. 
We order the general quasiparticle states and place the restrictions of $\mu<\nu$, $\mu^\prime<\nu^\prime$ in $C^{(K^\prime\pi^\prime)I}_{\mu\nu,\mu^\prime\nu^\prime}$  
without losing generality. 
These conditions are also applied to 
$C^{(K^\prime\pi^\prime)F}_{\mu\nu,\mu^\prime\nu^\prime}$.
If $K^\prime$ is equal to 0, Eqs.~(\ref{eq:vhatI}) and 
(\ref{eq:vhatF}) contain the same product of the creation operators twice; that is our choice of convention for simplicity of the programming of the code. 

Solving the QRPA equation, we obtain 
\begin{eqnarray}
&& O^{I\dagger}_{m^\prime} = \sum_{\mu < \nu}\left( X^{Im^\prime}_{\mu\nu} a^{I\dagger}_\mu a^{I\dagger}_\nu 
 - Y^{Im^\prime}_{-\mu-\nu} a^I_{-\nu} a^I_{-\mu} \right)~,\\
&& O^{I}_{m^\prime} = \sum_{\mu < \nu}\left( X^{Im^\prime\ast}_{\mu\nu} a^{I}_\nu a^{I}_\mu 
 - Y^{Im^\prime\ast}_{-\mu-\nu} a^{I\dagger}_{-\mu} a^{I\dagger}_{-\nu} \right)~,\\
&& O^{F\dagger}_{m^\prime} = \sum_{\mu < \nu}\left( X^{Fm^\prime}_{\mu\nu} a^{F\dagger}_\mu a^{F\dagger}_\nu 
 - Y^{Fm^\prime}_{-\mu-\nu} a^F_{-\nu} a^F_{-\mu} \right)~,\\
&& O^{F}_{m^\prime} = \sum_{\mu < \nu}\left( X^{Fm^\prime\ast}_{\mu\nu} a^{F}_\nu a^{F}_\mu 
 - Y^{Fm^\prime\ast}_{-\mu-\nu} a^{F\dagger}_{-\mu} a^{F\dagger}_{-\nu} \right)~,
\label{eq:O_QRPA}
\end{eqnarray}
where $j^z_\mu+j^z_\nu = K_{m^\prime}$ and $\pi_\mu \pi_\nu=\pi_{m^\prime}$.

We define matrices
\begin{eqnarray}
 &&C^{(K^\prime\pi^\prime)I} = \left( 
 \begin{array}{ccc}
  C^{(K^\prime\pi^\prime)I}_{11,-1-1} & \cdots & C^{(K^\prime\pi^\prime)I}_{11,-n-n^\prime} \\
   & \cdots & \\
  C^{(K^\prime\pi^\prime)I}_{nn^\prime,-1-1} & \cdots & C^{(K^\prime\pi^\prime)I}_{nn^\prime,-n-n^\prime}
 \end{array}
 \right)~,
\\
\nonumber\\
 &&X^{(K^\prime\pi^\prime)I} = \left( 
 \begin{array}{ccc}
  X^{I1}_{11} & \cdots & X^{IM}_{11} \\
    & \cdots & \\
  X^{I1}_{nn^\prime} & \cdots & X^{IM}_{nn^\prime}
 \end{array}
 \right)~,~\\
 &&Y^{(K^\prime\pi^\prime)I} = \left( 
 \begin{array}{ccc}
  Y^{I1}_{-1-1} & \cdots & Y^{IM}_{-1-1} \\
    & \cdots & \\
  Y^{I1}_{-n-n^\prime} & \cdots & Y^{IM}_{-n-n^\prime}
 \end{array}
 \right)~,
\end{eqnarray}
where the QRPA solutions of the $(K^\prime\pi^\prime)$ are used. Matrices $C^{(K^\prime\pi^\prime)F}$, $X^{(K^\prime\pi^\prime)F}$, and $Y^{(K^\prime\pi^\prime)F}$ are also introduced in the same manner.

Subsequently, we obtain $C^{(K^\prime\pi^\prime)I}$ and $C^{(K^\prime\pi^\prime)F}$ ignoring the exchange terms (the quasi-boson approximation 
\cite{Row68}), 
\begin{eqnarray}
 C^{(K^\prime\pi^\prime)I} = \frac{1}{1+\delta_{K0}}
\left( Y^{(K^\prime\pi^\prime)I} \frac{1}{X^{(K^\prime\pi^\prime)I}} \right)^\textrm{T}~, \nonumber \\
 C^{(K^\prime\pi^\prime)F} = \frac{1}{1+\delta_{K0}}
\left( Y^{(K^\prime\pi^\prime)F} \frac{1}{X^{(K^\prime\pi^\prime)F}} \right)^\textrm{T}~, \label{eq:CI_CF}
\end{eqnarray}
where the suffix T indicates the transpose of matrix, and it is assumed that $1/X^{(K^\prime\pi^\prime)I}$ and $1/X^{(K^\prime\pi^\prime)F}$ do not have a singularity.

The relation between the two HFB states can be written as, see e.g.~\cite{Rin80}, 
\begin{eqnarray}
|i\rangle = \frac{1}{{\cal N}_i} \exp\left[ \sum_{\mu\nu} D_{\mu\nu} a^{F\dagger}_\mu a^{F\dagger}_\nu \right] | f\rangle~, \label{eq:i-f} \\
{\cal N}_i = \frac{1}{\langle f|i\rangle} = \sqrt{\det(I+D^\dagger D)}~, \label{eq:N_HFB} \\
D = \left(
\begin{array}{ccc}
 D_{1-1} & \cdots & D_{1-n} \\
   & \cdots & \\
 D_{n-1} & \cdots & D_{n-n}
\end{array}
\right)~.
\end{eqnarray}
Here, $I$ denotes the unit matrix with the size of matrix $D$, and
$D_{\mu\nu}$ is not equal to zero only for those $\mu$ and $\nu$ satisfying $j^z_\mu+j^z_\nu = 0 $ and 
$\pi_\mu \pi_\nu = +$. We place the restriction of $j^z_\mu>0$ in Eq.~(\ref{eq:i-f}).
\noindent
The unitary transformation from the basis $\{a^{F\dagger}_\mu, a^{F}_{-\mu} \}$ to the basis $\{ a^{I\dagger}_\mu, a^{I}_{\mu} \}$ is given by
\begin{eqnarray}
a^{I\dagger}_\mu = \sum_{\mu^\prime} \left(
 T^{IF1}_{\mu\mu^\prime} a^{F\dagger}_{\mu^\prime} + T^{IF2}_{\mu-\mu^\prime} a^F_{-\mu^\prime} \right)~, \\
a^I_\mu = \sum_{\mu^\prime} \left(
 T^{IF1\ast}_{\mu\mu^\prime} a^F_{\mu^\prime} + T^{IF2*}_{\mu-\mu^\prime} a^{F\dagger}_{-\mu^\prime} \right)~, \label{eq:transformation_qp}
\end{eqnarray}
with $j^z_\mu=j^z_{\mu^\prime}$ and $\pi_\mu=\pi_{\mu^\prime}$. 
The matrix elements of the unitary transformation can be calculated as 
\begin{eqnarray}
T^{IF1}_{\mu\mu^\prime}&=&
\int d^3\bm{r}\sum_\sigma \biglb( U^{F\ast}_{\mu^\prime}(\bm{r},\sigma) U^I_\mu(\bm{r},\sigma)
\nonumber \\
&&+V^{F\ast}_{\mu^\prime}(\bm{r},\sigma) V^I_\mu(\bm{r},\sigma) \bigrb)~, \label{eq:TF1} \\
T^{IF2}_{\mu -\mu^\prime}&=&
\int d^3\bm{r} \sum_\sigma \biglb( V^F_{-\mu^\prime}(\bm{r},\sigma) U^I_\mu(\bm{r},\sigma)
\nonumber \\
&&+U^F_{-\mu^\prime}(\bm{r},\sigma) V^I_\mu(\bm{r},\sigma) \bigrb)~, 
\label{eq:TF2}
\end{eqnarray}
by using the wave functions of the general quasiparticle, see e.g.~\cite{Dob84}, 
\begin{equation}
\left(
\begin{array}{l}
U^I_\mu(\bm{r},\sigma)\\
V^I_{\mu}(\bm{r},-\sigma) \label{eq:qp_wf}
\end{array}
\right),
\end{equation}
and those associated with the state $F$, where $\sigma=\pm 1/2$ is the $z$-component of the spin. 
$D_{\mu -\nu}$ is obtained from
\begin{equation}
D = -\left( \frac{1}{T^{IF1}}T^{IF2}\right)^\ast~, \label{eq:D}
\end{equation}
where the matrices used are defined as
\begin{eqnarray}
 T^{IF1} &=& \left( 
\begin{array}{ccc}
 T^{IF1}_{11} & \cdots & T^{IF1}_{1n} \\
   & \cdots & \\
 T^{IF1}_{n1} & \cdots & T^{IF1}_{nn}
\end{array}
\right)~,\\
\nonumber\\
 T^{IF2} &=& \left( 
\begin{array}{ccc}
 T^{IF2}_{1-1} & \cdots & T^{IF2}_{1-n} \\
   & \cdots & \\
 T^{IF2}_{n-1} & \cdots & T^{IF2}_{n-n}
\end{array}
\right)~,
\end{eqnarray}
and it is assumed that $1/T^{IF1}$ does not have a singularity.

\begin{widetext}
Now, we expand and truncate the overlap matrix element with respect to $\hat{v}^{(K^\prime\pi^\prime)}_F$ and 
$\hat{v}^{(K^\prime\pi^\prime)}_I$
\begin{eqnarray}
 \langle F | O^F_m O^{I\dagger}_{m^\prime} |I\rangle &=& 
\frac{1}{{\cal N}^\prime_I{\cal N}^\prime_F}
\langle f | \prod_{K_1\pi_1}\exp\left[\hat{v}^{(K_1\pi_1)\dagger}_F\right] O^F_m O^{I\dagger}_{m^\prime} \prod_{K_2\pi_2}\exp\left[ \hat{v}^{(K_2\pi_2)}_I \right] | i \rangle \nonumber\\
&\simeq& 
 {\cal M} 
 \Bigg\{ 
 \langle f | O^F_m O^{I\dagger}_{m^\prime} | i \rangle + 
 \sum_{K_1 \pi_1} \left( \langle f | \hat{v}^{(K_1\pi_1)\dagger}_F O^F_m O^{I\dagger}_{m^\prime} | i \rangle 
+ \langle f | O^F_m O^{I\dagger}_{m^\prime} \hat{v}^{(K_1\pi_1)}_I | i \rangle \right) 
\nonumber \\
 &&+\sum_{K_1\pi_1} \langle f | \hat{v}^{(K_1\pi_1)\dagger}_F O^F_m O^{I\dagger}_{m^\prime} \hat{v}^{(K_1\pi_1)}_I | i \rangle
 \Bigg\},
 \label{eq:fooi}
\end{eqnarray}
\begin{equation}
{\cal M} = \frac{1}{ {\cal N}_I {\cal N}_F }, \label{eq:M}
\end{equation}
\begin{eqnarray}
 &&{\cal N}_I \simeq \sqrt{ 1+\sum_{K_1\pi_1} \left\{ \langle i | \hat{v}^{(K_1\pi_1)\dagger}_I 
\hat{v}^{(K_1\pi_1)}_I | i \rangle 
+\frac{1}{4} \langle i | \left( \hat{v}^{(K_1\pi_1)\dagger}_I \right)^2 \left( \hat{v}^{(K_1\pi_1)}_I \right)^2 | i \rangle 
\right\} }~, \label{eq:NI_QRPA} \\
 &&{\cal N}_F \simeq \sqrt{ 1+\sum_{K_1\pi_1} \left\{ \langle f | \hat{v}^{(K_1\pi_1)\dagger}_F 
\hat{v}^{(K_1\pi_1)}_F | f \rangle 
+\frac{1}{4}\langle f | \left( \hat{v}^{(K_1\pi_1)\dagger}_F \right)^2 \left( \hat{v}^{(K_1\pi_1)}_F \right)^2 | f \rangle
\right\} }~, \label{eq:NF_QRPA}
\end{eqnarray}
\begin{equation}
\langle i|\hat{v}^{(K_1\pi_1)\dagger}_I\hat{v}^{(K_1\pi_1)}_I |i\rangle = 
\langle i|\hat{v}^{(K_1\pi_1)\dagger}_I\hat{v}^{(K_1\pi_1)}_I |i\rangle_{\textrm{boson}} + 
\langle i|\hat{v}^{(K_1\pi_1)\dagger}_I\hat{v}^{(K_1\pi_1)}_I |i\rangle_{\textrm{exch}},
\end{equation}
\begin{equation}
\langle i|\hat{v}^{(K_1\pi_1)\dagger}_I\hat{v}^{(K_1\pi_1)}_I |i\rangle_\textrm{boson}
= (1+\delta_{K_1 0})\textrm{Tr}\left(C^{(K_1\pi_1)I}C^{(K_1\pi_1)I\dagger}\right), 
\end{equation}
\begin{eqnarray}
\lefteqn{\langle i | \hat{v}^{(K_1\pi_1)\dagger}_I \hat{v}^{(K_1\pi_1)}_I | i \rangle_\textrm{exch}} \nonumber \\
&=& (1+\delta_{K_1 0})\sum_{\mu\nu}\sum_{\mu^\prime\nu^\prime}C^{(K_1\pi_1)I\ast}_{\mu\nu,\mu^\prime\nu^\prime} 
\Big(  -C^{(K_1\pi_1)I}_{\mu^\prime\mu,\nu^\prime\nu}
+C^{(K_1\pi_1)I}_{\mu^\prime\mu,\nu\nu^\prime}
-C^{(K_1\pi_1)I}_{\nu^\prime\mu,\nu\mu^\prime}
+C^{(K_1\pi_1)I}_{\nu^\prime\mu,\mu^\prime\nu} \nonumber \\
&&+C^{(K_1\pi_1)I}_{\mu\mu^\prime,\nu^\prime\nu}
-C^{(K_1\pi_1)I}_{\mu\mu^\prime,\nu\nu^\prime}
+C^{(K_1\pi_1)I}_{\mu\nu^\prime,\nu\mu^\prime}
-C^{(K_1\pi_1)I}_{\mu\nu^\prime,\mu^\prime\nu}
\Big), \label{eq:exc_v2_nrm_i}
\end{eqnarray}
\begin{equation}
\langle f|\hat{v}^{(K_1\pi_1)\dagger}_F\hat{v}^{(K_1\pi_1)}_F |f\rangle = 
\langle f|\hat{v}^{(K_1\pi_1)\dagger}_F\hat{v}^{(K_1\pi_1)}_F |f\rangle_{\textrm{boson}} + 
\langle f|\hat{v}^{(K_1\pi_1)\dagger}_F\hat{v}^{(K_1\pi_1)}_F |f\rangle_{\textrm{exch}},
\end{equation}
\begin{equation}
\langle f|\hat{v}^{(K_1\pi_1)\dagger}_F\hat{v}^{(K_1\pi_1)}_F |f\rangle_\textrm{boson}
= (1+\delta_{K_1 0})\textrm{Tr}\left(C^{(K_1\pi_1)F}C^{(K_1\pi_1)F\dagger}\right), 
\end{equation}
\begin{eqnarray}
\lefteqn{\langle f | \hat{v}^{(K_1\pi_1)\dagger}_F \hat{v}^{(K_1\pi_1)}_F | f \rangle_\textrm{exch}} \nonumber \\
&=& (1+\delta_{K_1 0})\sum_{\mu\nu}\sum_{\mu^\prime\nu^\prime}C^{(K_1\pi_1)F\ast}_{\mu\nu,\mu^\prime\nu^\prime} 
\Big(  -C^{(K_1\pi_1)F}_{\mu^\prime\mu,\nu^\prime\nu}
+C^{(K_1\pi_1)F}_{\mu^\prime\mu,\nu\nu^\prime}
-C^{(K_1\pi_1)F}_{\nu^\prime\mu,\nu\mu^\prime}
+C^{(K_1\pi_1)F}_{\nu^\prime\mu,\mu^\prime\nu} \nonumber \\
&&+C^{(K_1\pi_1)F}_{\mu\mu^\prime,\nu^\prime\nu}
-C^{(K_1\pi_1)F}_{\mu\mu^\prime,\nu\nu^\prime}
+C^{(K_1\pi_1)F}_{\mu\nu^\prime,\nu\mu^\prime}
-C^{(K_1\pi_1)F}_{\mu\nu^\prime,\mu^\prime\nu}
\Big), \label{eq:exc_v2_nrm_f}
\end{eqnarray}
The fourth-order terms in Eqs.~(\ref{eq:NI_QRPA}) and (\ref{eq:NF_QRPA}) are approximated by the following quasi-boson terms
\begin{eqnarray}
\lefteqn{\frac{1}{4}\sum_{K_1\pi_1}\langle i | \left( \hat{v}^{(K_1\pi_1)\dagger}_I \right)^2 \left( \hat{v}^{(K_1\pi_1)}_I \right)^2 | i \rangle}
\nonumber\\
&\simeq& \frac{1}{4}\sum_{K_1\pi_1}\bigg\{ (2+6\delta_{K_1 0}) \left( \textrm{Tr} (C^{(K_1\pi_1)I} C^{(K_1\pi_1)I\dagger}) \right)^2
+ (2+14\delta_{K_1 0})\textrm{Tr}(C^{(K_1\pi_1)I} C^{(K_1\pi_1)I\dagger})^2 \bigg\}, \label{eq:fourth_norm_i}\\
\lefteqn{\frac{1}{4}\sum_{K_1\pi_1}\langle f | \left( \hat{v}^{(K_1\pi_1)\dagger}_F \right)^2 \left( \hat{v}^{(K_1\pi_1)}_F \right)^2 | f \rangle}
\nonumber\\
&\simeq& \frac{1}{4}\sum_{K_1\pi_1}\bigg\{ (2+6\delta_{K_1 0}) \left( \textrm{Tr} (C^{(K_1\pi_1)F} C^{(K_1\pi_1)F\dagger}) \right)^2
+ (2+14\delta_{K_1 0})\textrm{Tr}(C^{(K_1\pi_1)F} C^{(K_1\pi_1)F\dagger})^2 \bigg\}.
\label{eq:fourth_norm}
\end{eqnarray}
\end{widetext}
We test up to the second-order terms 
\begin{equation}
\langle f | \hat{v}^{(K_1\pi_1)\dagger}_F O^F_m O^{I\dagger}_{m^\prime} \hat{v}^{(K_1 \pi_1)}_I | i \rangle,
\label{eq:2nd_overlap}
\end{equation}
with respect to $\hat{v}_F^{(K_1\pi_1)}$ and $\hat{v}_I^{(K_2\pi_2)}$ but only with $(K_1\pi_1)= (K_2\pi_2)$ in Eq.~(\ref{eq:fooi})\footnote{It was found in the numerical calculation shown in Sec.~\ref{sec:result} that 
the summation of Eq.~(\ref{eq:2nd_overlap}) with respect to $(K_1\pi_1)$ was negligible, 
and thus, we did not calculate the terms for which $(K_1\pi_1)\neq (K_2\pi_2)$.}. 
The terms up to the fourth-order are included in the normalization factors ${\cal N}_I$ and ${\cal N}_F$, because its convergence is slow 
with respect to the $\hat{v}$-expansion compared to the un-normalized overlap matrix elements, that is, Eq.~(\ref{eq:fooi}) without ${\cal M}$. 
The reason for this difference is that the bra and ket HFB ground states are identical in the normalization factors, while these states are quite different 
around the Fermi surface in the un-normalized overlap. Due to this difference, high-energy excitations leaving the configuration around the Fermi surface intact do not significantly contribute to the un-normalized overlap matrix. 
Hence, the un-normalized overlap has less of the major terms than the normalization factors, 
and $\hat{v}^{(K_1 \pi_1)}_F$ $(\hat{v}^{(K_1 \pi_1)}_I)$ in the un-normalized overlap has a smaller effect than in the normalization factors 
(see the numerical results in Sec.~\ref{sec:result}). 

\begin{widetext}

The first term of Eq.~(\ref{eq:fooi}) is obtained
\begin{eqnarray}
 {\cal M}
 \langle f | O^F_m O^{I\dagger}_{m^\prime} | i \rangle &=& 
 {\cal M}
 \sum_{\mu<\nu} X^{Fm\ast}_{\mu\nu} \sum_{\mu^\prime<\nu^\prime} X^{Im^\prime}_{\mu^\prime\nu^\prime}
  \langle f | a^F_\nu a^F_\mu a^{I\dagger}_{\mu^\prime} a^{I\dagger}_{\nu^\prime}| i \rangle~.
\label{eq:ft}
\end{eqnarray}
The second term of Eq.~(\ref{eq:fooi}) reads
\begin{eqnarray}
 &{\cal M}&
\sum_{K_1\pi_1}\lefteqn{\langle f | \hat{v}^{(K_1\pi_1)\dagger}_F O^F_m O^{I\dagger}_{m^\prime} | i \rangle }\nonumber\\
&&={\cal M}\!\!
 \sum_{K_1\pi_1}\sum_{\mu\nu\mu^\prime\nu^\prime}\sum_{\mu_1<\nu_1}\sum_{\mu_2<\nu_2}
 C^{(K_1\pi_1)F\ast}_{\mu\nu,\mu^\prime\nu^\prime} X^{Fm\ast}_{\mu_1\nu_1} X^{Im^\prime}_{\mu_2\nu_2}
 \langle f | a^F_{\nu^\prime} a^F_{\mu^\prime} a^F_{\nu} a^F_{\mu} a^F_{\nu_1} a^F_{\mu_1} a^{I\dagger}_{\mu_2} a^{I\dagger}_{\nu_2}|i\rangle \nonumber \\
&&\hspace{10pt}-{\cal M} 
\sum_{K_1\pi_1}\sum_{\mu\nu}\sum_{\mu_1<\nu_1}\sum_{\mu_2<\nu_2} Y^{Fm\ast}_{-\mu_1-\nu_1} X^{Im^\prime}_{\mu_2\nu_2}
\Big\{ C^{(K_1\pi_1)F\ast}_{-\nu_1-\mu_1,\mu\nu} -C^{(K_1 \pi_1)F\ast}_{-\mu_1-\nu_1,\mu\nu} 
\nonumber \\
 &&\hspace{10pt}+C^{(K_1\pi_1)F\ast}_{\mu\nu,-\nu_1-\mu_1} - C^{(K_1\pi_1)F\ast}_{\mu\nu,-\mu_1-\nu_1}
 + C^{(K_1\pi_1)F\ast}_{-\nu_1\nu,-\mu_1\mu} - C^{(K_1\pi_1)F\ast}_{-\mu_1\nu,-\nu_1\mu} 
  -C^{(K_1\pi_1)F\ast}_{-\nu_1\nu,\mu-\mu_1} + C^{(K_1\pi_1)F\ast}_{-\mu_1\nu,\mu-\nu_1} \nonumber\\
 &&\hspace{10pt}+C^{(K_1\pi_1)F\ast}_{\mu-\nu_1,-\mu_1\nu} - C^{(K_1\pi_1)F\ast}_{\mu-\mu_1,-\nu_1\nu} 
  -C^{(K_1\pi_1)F\ast}_{\mu-\nu_1,\nu-\mu_1} + C^{(K_1\pi_1)F\ast}_{\mu-\mu_1,\nu-\nu_1} \Big\}
\langle f | a^F_\mu a^F_\nu a^{I\dagger}_{\mu_2} a^{I\dagger}_{\nu_2} | i \rangle~. 
\label{eq:st}
\end{eqnarray}
The third term of Eq.~(\ref{eq:fooi}) is given by
\begin{eqnarray}
 &{\cal M}&\!\!
\sum_{K_1\pi_1} 
\lefteqn{ \langle f | O^F_m O^{I\dagger}_{m^\prime} \hat{v}^{(K_1\pi_1)}_I | i \rangle } \nonumber\\
&&={\cal M}
\sum_{K_1\pi_1}\sum_{\mu<\nu}\sum_{\mu^\prime<\nu^\prime}\sum_{\mu_1\nu_1}\sum_{\mu_2\nu_2}
X^{Fm\ast}_{\mu\nu} X^{Im^\prime}_{\mu^\prime\nu^\prime} C^{(K_1\pi_1)I}_{\mu_1\nu_1,\mu_2\nu_2}
\langle f | a^F_\nu a^F_\mu a^{I\dagger}_{\mu^\prime} a^{I\dagger}_{\nu^\prime} a^{I\dagger}_{\mu_1} a^{I\dagger}_{\nu_1} a^{I\dagger}_{\mu_2} a^{I\dagger}_{\nu_2} | i \rangle \nonumber\\
&&\hspace{10pt}-{\cal M} 
\sum_{K_1\pi_1}\sum_{\mu<\nu} \sum_{\mu^\prime<\nu^\prime} \sum_{\mu_1 \mu_2} X^{Fm\ast}_{\mu\nu} Y^{Im^\prime}_{-\mu^\prime-\nu^\prime}
\Big\{
-C^{(K_1\pi_1)I}_{\mu_1\mu_2,-\nu^\prime-\mu^\prime} + C^{(K_1\pi_1)I}_{\mu_1\mu_2,-\mu^\prime-\nu^\prime} \nonumber \\
&&\hspace{10pt}-C^{(K_1\pi_1)I}_{-\nu^\prime-\mu^\prime,\mu_1\mu_2} + C^{(K_1\pi_1)I}_{-\mu^\prime-\nu^\prime,\mu_1\mu_2}
-C^{(K_1\pi_1)I}_{\mu_1-\mu^\prime,\mu_2-\nu^\prime} + C^{(K_1\pi_1)I}_{\mu_1-\nu^\prime,\mu_2-\mu^\prime} 
+C^{(K_1\pi_1)I}_{\mu_1-\mu^\prime,-\nu^\prime\mu_2} - C^{(K_1\pi_1)I}_{\mu_1-\nu^\prime,-\mu^\prime\mu_2} \nonumber \\
&&\hspace{10pt}+C^{(K_1\pi_1)I}_{-\mu^\prime\mu_1,\mu_2-\nu^\prime} - C^{(K_1\pi_1)I}_{-\nu^\prime\mu_1,\mu_2-\mu^\prime}
-C^{(K_1\pi_1)I}_{-\mu^\prime\mu_1,-\nu^\prime\mu_2} + C^{(K_1\pi_1)I}_{-\nu^\prime\mu_1,-\mu^\prime\mu_2} 
\Big\} \langle f | a^F_\nu a^F_\mu a^{I\dagger}_{\mu_1} a^{I\dagger}_{\mu_2} | i \rangle~.
\label{eq:tt}
\end{eqnarray}
Further, we can write the fourth term of Eq.~(\ref{eq:fooi}) as follows:
\begin{equation}
{\cal M}
\sum_{K_1\pi_1}\langle f | \hat{v}^{(K_1\pi_1)\dagger}_F O^F_m O^{I\dagger}_{m^\prime} \hat{v}^{K_1\pi_1}_I | i \rangle
= F^1_{mm^\prime} + F^2_{mm^\prime} + F^3_{mm^\prime} + F^4_{mm^\prime}~,
\label{eq:fot}
\end{equation}
\begin{eqnarray}
F^1_{mm^\prime} &=& 
{\cal M}
\sum_{K_1\pi_1}\sum_{\mu\nu\mu^\prime\nu^\prime} \sum_{{\mu_3}{\nu_3}{\mu_4}{\nu_4}}
C^{(K_1\pi_1)F\ast}_{\mu\nu,\mu^\prime\nu^\prime} 
C^{(K_1\pi_1)I}_{ {\mu_3}{\nu_3},{\mu_4}{\nu_4} }
\sum_{ \mu_1<\nu_1 } \sum_{ \mu_2<\nu_2 }
X^{Fm}_{\mu_1\nu_1} X^{Im^\prime}_{\mu_2\nu_2}
\nonumber\\
&&\times\langle f |
a^F_{\nu^\prime} a^F_{\mu^\prime} a^F_\nu a^F_\mu a^F_{\nu_1} a^F_{\mu_1} a^{I\dagger}_{\mu_2} a^{I\dagger}_{\nu_2} a^{I\dagger}_{{\mu_3}} a^{I\dagger}_{{\nu_3}} a^{I\dagger}_{{\mu_4}} a^{I\dagger}_{{\nu_4}}
| i \rangle ~,\label{eq:fot1} \\
F^2_{mm^\prime} &=&
{\cal M}
\sum_{K_1\pi_1}\sum_{{\mu_3}{\nu_3} }
\sum_{ \mu_1\nu_1 }
X^{Fm}_{\mu_1\nu_1} C^{(K_1\pi_1)X2}_{\mu_1\nu_1,{\mu_3}{\nu_3} }
\sum_{\mu_2<\nu_2} Y^{Im^\prime}_{-\mu_2-\nu_2}
\Big( -C^{(K_1\pi_1)I}_{ {\mu_3}{\nu_3}, -\mu_2-\nu_2 }
+C^{(K_1\pi_1)I}_{ {\mu_3}-\mu_2, {\nu_3}-\nu_2 } \nonumber \\
&&
-C^{(K_1\pi_1)I}_{ {\mu_3}-\mu_2, -\nu_2{\nu_3} }
+C^{(K_1\pi_1)I}_{-\mu_2{\nu_3}, {\mu_3}-\nu_2 }
-C^{(K_1\pi_1)I}_{-\mu_2{\nu_3},-\nu_2{\mu_3} }
-C^{(K_1\pi_1)I}_{{\mu_3}-\nu_2, {\nu_3}-\mu_2 }
+C^{(K_1\pi_1)I}_{{\mu_3}-\nu_2, -\mu_2{\nu_3} }
-C^{(K_1\pi_1)I}_{-\nu_2{\nu_3}, {\mu_3}-\mu_2 } \nonumber \\
&&
+C^{(K_1\pi_1)I}_{-\nu_2{\nu_3},-\mu_2{\mu_3} } 
-C^{(K_1\pi_1)I}_{-\mu_2-\nu_2,{\mu_3}{\nu_3}}
\Big)~, \label{eq:fot2}
\end{eqnarray}
\begin{equation}
C^{(K_1\pi_1)X2}_{\mu_1\nu_1,{\mu_3}{\nu_3} } = 
\sum_{\mu\nu} \sum_{\mu^\prime\nu^\prime} C^{(K_1\pi_1)F\ast}_{ \mu\nu, \mu^\prime\nu^\prime }
\langle f |
a^F_{\nu^\prime} a^F_{\mu^\prime} a^F_\nu a^F_\mu a^F_{\nu_1} a^F_{\mu_1} 
a^{I\dagger}_{{\mu_3}} a^{I\dagger}_{{\nu_3}}
| i \rangle~, \label{eq:fot2x2}
\end{equation}
\begin{eqnarray}
F^3_{mm^\prime} &=&
-{\cal M} \sum_{K_1\pi_1}\sum_{\mu\nu} \sum_{\mu_2<\nu_2}
X^{Im^\prime}_{\mu_2\nu_2} C^{(K_1\pi_1)X3}_{\mu\nu,\mu_2\nu_2}
\sum_{\mu^\prime\nu^\prime} Y^{Fm}_{\mu^\prime\nu^\prime}
\Big( C^{(K_1\pi_1)F\ast}_{\mu\nu,\mu^\prime\nu^\prime}
-C^{(K_1\pi_1)F\ast}_{ \mu\mu^\prime, \nu\nu^\prime } \nonumber \\
&&+C^{(K_1\pi_1)F\ast}_{ \mu\mu^\prime, \nu^\prime\nu }
-C^{(K_1\pi_1)F\ast}_{ \mu^\prime\nu, \mu\nu^\prime }
+C^{(K_1\pi_1)F\ast}_{ \mu^\prime\nu, \nu^\prime\mu }
+C^{(K_1\pi_1)F\ast}_{ \mu\nu^\prime, \nu\mu^\prime }
-C^{(K_1\pi_1)F}_{ \mu\nu^\prime, \mu^\prime\nu }
+C^{(K_1\pi_1)F\ast}_{ \nu^\prime\nu, \mu\mu^\prime } \nonumber \\
&&
-C^{(K_1\pi_1)F\ast}_{\nu^\prime\nu, \mu^\prime\mu} 
+C^{(K_1\pi_1)F\ast}_{\mu^\prime\nu^\prime,\mu\nu}
\Big)~, \label{eq:fot3}
\end{eqnarray}
\begin{equation}
C^{(K_1\pi_1)X3}_{ \mu\nu, \mu_2\nu_2 } = \sum_{{\mu_3}{\nu_3}}
\sum_{ {\mu_4} {\nu_4} }
C^{(K_1\pi_1)I}_{ {\mu_3}{\nu_3}, {\mu_4}{\nu_4}}
\langle f | 
a^F_\nu a^F_\mu a^{I\dagger}_{\mu_2} a^{I\dagger}_{\nu_2} a^{I\dagger}_{{\mu_3}} a^{I\dagger}_{{\nu_3}} a^{I\dagger}_{{\mu_4}} a^{I\dagger}_{{\nu_4}}
|i\rangle~,
\label{fot3x3}
\end{equation}
\begin{eqnarray}
F^4_{mm^\prime}&=&
{\cal M} \sum_{K_1\pi_1}\sum_{\mu^\prime\nu^\prime} \sum_{{\mu_4}{\nu_4}}
\sum_{\mu\nu} \sum_{ {\mu_3} {\nu_3} }
\Big(
 C^{(K_1\pi_1)F\ast}_{ \mu\nu, -\mu^\prime-\nu^\prime }
-C^{(K_1\pi_1)F\ast}_{ \mu-\mu^\prime, \nu-\nu^\prime }
+C^{(K_1\pi_1)F\ast}_{ \mu-\mu^\prime, -\nu^\prime\nu } \nonumber \\
&&
+C^{(K_1\pi_1)F\ast}_{-\mu^\prime\mu, \nu-\nu^\prime }
-C^{(K_1\pi_1)F\ast}_{ -\mu^\prime\mu, -\nu^\prime\nu }
+C^{(K_1\pi_1)F\ast}_{ \mu-\nu^\prime, \nu-\mu^\prime }
-C^{(K_1\pi_1)F\ast}_{ \mu-\nu^\prime, -\mu^\prime\nu }
-C^{(K_1\pi_1)F\ast}_{ -\nu^\prime\mu, \nu-\mu^\prime }
+C^{(K_1\pi_1)F\ast}_{-\nu^\prime\mu, -\mu^\prime\nu }
+C^{(K_1\pi_1)F\ast}_{-\mu^\prime-\nu^\prime,\mu\nu}
\Big) \nonumber \\
&&
\times\langle f | a^F_\nu a^F_\mu a^{I\dagger}_{{\mu_3}} a^{I\dagger}_{{\nu_3}} | i \rangle
\Big(
 C^{(K_1\pi_1)I}_{ {\mu_3} {\nu_3}, -{\mu_4} -{\nu_4} }
-C^{(K_1\pi_1)I}_{ {\mu_3} -{\mu_4}, {\nu_3} -{\nu_4} }
+C^{(K_1\pi_1)I}_{ {\mu_3} -{\mu_4}, -{\nu_4} {\nu_3} }
+C^{(K_1\pi_1)I}_{ -{\mu_4} {\mu_3}, {\nu_3} -{\nu_4} }
-C^{(K_1\pi_1)I}_{ -{\mu_4} {\mu_3}, -{\nu_4} {\nu_3} }
\nonumber \\
&&
+C^{(K_1\pi_1)I}_{ {\mu_3} -{\nu_4}, {\nu_3} -{\mu_4} }
-C^{(K_1\pi_1)I}_{ {\mu_3} -{\nu_4}, -{\mu_4} {\nu_3} } 
-C^{(K_1\pi_1)I}_{ -{\nu_4} {\mu_3}, {\nu_3} -{\mu_4} }
+C^{(K_1\pi_1)I}_{ -{\nu_4} {\mu_3}, -{\mu_4} {\nu_3} }
+C^{(K_1\pi_1)I}_{-{\mu_4}-{\nu_4}, {\mu_3}{\nu_3} }
\Big)
Y^{Fm}_{-\mu^\prime-\nu^\prime} Y^{Im^\prime}_{ -{\mu_4} -{\nu_4} }~.
\label{eq:fot4}
\end{eqnarray}

The generalized expectation value of the multiple fermion operators can be calculated by using the generalized Wick's theorem \cite{Bal69}. 
In particular, that of  four operators can be written explicitly
\begin{equation}
\langle f | a^F_{\nu^\prime} a^F_{\mu^\prime} a^{I\dagger}_{\mu_2} a^{I\dagger}_{\nu_2} | i \rangle
= \frac{1}{\langle f | i \rangle} \Big( 
\langle f | a^F_{\nu^\prime} a^F_{\mu^\prime} | i \rangle \langle f | a^{I\dagger}_{\mu_2} a^{I\dagger}_{\nu_2} | i \rangle
- \langle f | a^F_{\nu^\prime} a^{I\dagger}_{\mu_2} | i \rangle \langle f | a^F_{\mu^\prime} a^{I\dagger}_{\nu_2} | i \rangle
+\langle f | a^F_{\nu^\prime} a^{I\dagger}_{\nu_2} | i \rangle \langle f |a^F_{\mu^\prime} a^{I\dagger}_{\mu_2} | i \rangle
\Big)~, \label{eq:wick4}
\end{equation} 
with the following contractions
\begin{eqnarray}
&&\langle f | a^F_\mu a^F_{-\nu} | i \rangle = 
\left\{
\begin{array}{l}
\displaystyle{
-\frac{1}{{\cal N}_i}D_{\mu-\nu}~,~j^z_\mu>0~,
}\\
\\
-\langle f | a^F_{-\nu} a^F_\mu | i \rangle~,~j^z_\mu<0~,
\end{array}
\right. \label{eq:contraction1} \\
\nonumber\\
&&\langle f | a^{I\dagger}_\mu a^{I\dagger}_{-\nu} | i \rangle = 
\left\{
\begin{array}{l}
\displaystyle{
\sum_{\mu^\prime}T^{IF2}_{\mu-\mu^\prime} T^{IF1}_{\nu\mu^\prime} t_\nu t^\ast_{\mu^\prime}
\frac{1}{{\cal N}_i} 
-\sum_{\mu^\prime} T^{IF2}_{\mu -\mu^\prime} \sum_{\nu^\prime}T^{IF2\ast}_{\nu -\nu^\prime}
t_\nu t^\ast_{-\nu^\prime}
\langle f | a^F_{\nu^\prime} a^F_{-\mu^\prime} | i \rangle~,~ j^z_\mu > 0 ~,
}\\
\\
-\langle f | a^I_{-\nu} a^{I\dagger}_{\mu} | i \rangle~,~ j^z_\mu < 0~,
\end{array}
\right. \label{eq:contraction2}\\
\nonumber\\
&&\langle f | a^F_\mu a^{I\dagger}_\nu | i \rangle = 
\left\{
\begin{array}{l}
\displaystyle{
\frac{1}{{\cal N}_i}T^{IF1}_{\nu\mu} 
-\frac{1}{{\cal N}_i}\sum_{\nu^\prime} T^{IF2}_{\nu -\nu^\prime} D_{\mu -\nu^\prime}~,~j^z_\mu > 0~,
}\\
\\
t_{-\nu} t^\ast_{-\mu}\langle f | a^F_{-\mu} a^{I\dagger}_{-\nu} | i \rangle^\ast~,~j^z_\mu < 0~.\\
\end{array}
\right. \label{eq:contraction3}
\end{eqnarray}
\end{widetext}
\noindent
Here, $t_\mu$ is a phase arising from the time reversal of a fermion state as
\begin{equation}
\hat{T}|a\rangle=t^\ast_a|-a\rangle~,
\end{equation}
where $\hat{T}$ denotes the time-reversal operator \cite{Mes62}.

\section{\label{sec:analytical} Analytical properties of overlap matrix}
\subsection{\label{subsec:simple} Simple model}
\begin{figure}[h]
\includegraphics[width=8cm]{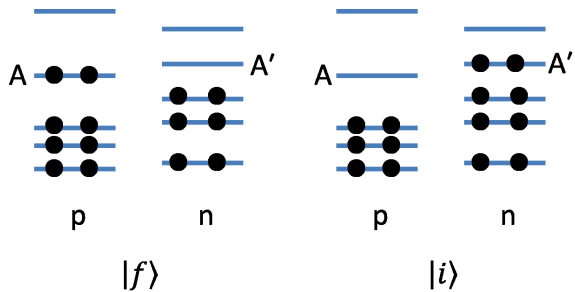}
\caption{ \label{fig:simple_model_1} (Color online) Simple model of final and initial states of $0\nu\beta\beta$ decay. The doubly-degenerated levels are in the relation of the time reversal. Levels A and A$^\prime$ are referred to in text. The notation p (n) denotes protons (neutrons). }
\end{figure}

Let us discuss the simple model shown in Fig.~\ref{fig:simple_model_1} for investigating the analytical properties of the overlap matrix elements of the QRPA states. A single-particle basis is shared by the initial and final states of the  
$0\nu\beta\beta$ decay with no pairing field. 
In this model, we assume that $|F \rangle = |f \rangle$ and $|I \rangle = |i \rangle$. When the QRPA is applied to this model, only three types of excitations are possible: two-particle addition, removal, or one-particle-one-hole excitation. Then, only two overlap matrix elements 
of the excited states are finite; one is the excited state with both levels A and A$^\prime$ (see Fig.~\ref{fig:simple_model_1}) occupied, and another is 
the excited state with none of those levels occupied. All the other overlap matrix elements vanish.
This feature can be quantified by the following measure 
\begin{equation}
{\cal S} = \textrm{Tr}( G^\dagger G)/\textrm{dim}(G), \label{eq:S}
\end{equation}
where $G$ denotes the overlap matrix, and dim($G$) denotes the dimension of matrix $G$. 
If $G$ is unitary, then, of course, ${\cal S}$ is equal to 1. The ${\cal S}$ value of our simple model is 
$2/\textrm{dim}{\cal S}$, which is of order of $10^{-4}$ or smaller with the dimension of the realistic calculations. Therefore, the overlap matrix discussed is highly non-unitary. One of its implications is that $O^F_m$ cannot be represented by a linear combination of $O^I_{m^\prime}$ and 
$O^{I\dagger}_{m^\prime}$. This is also seen from the nature of  the Bogoliubov transformation. Since $a^{F\dagger}_\mu$ and $a^{F}_\mu$ are represented by a linear combination of $a^{I\dagger}_{\mu^\prime}$ 
and $a^{I}_{\mu^\prime}$, $O^F_m$ includes the bilinear term $a^{I\dagger}_\mu a^{I}_{\mu^\prime}$. 
The appearance of this term is certain, because the two nuclei considered have different configurations.   
And, $a^{I\dagger}_\mu a^{I}_{\mu^\prime}$ is bilinear with respect to $O^{I\dagger}_{m^\prime}$ and $ O^I_{m^{\prime\prime}}$ according to the boson-expansion theories \cite{Kle91,Rin80}. 

The above argument using ${\cal S}$ indicates that the overlap matrix of the QRPA is not close to the one obtained from the exact many-body states  at all; the exact one has the absolute value of every diagonal matrix element equal to 1.
We need to recall the nuclear matrix element [Eq.~(\ref{eq:double_GT_our_original})] and Eq.~(\ref{eq:forbidden_2particle_transfer}) in order to understand the implication of this mathematical property of the overlap matrix. 
The inclusion of higher-order excited states such as
$ O^{I\dagger}_{m_1} O^{I\dagger}_{m_2}|I\rangle $
is necessary for having many diagonal overlap matrix elements close to 1; however, these states do not contribute to the two-particle transfer matrix elements.
Therefore, the QRPA has the possibility of being an efficient approximation to the nuclear matrix elements  irrespective of  the deviation of the overlap matrix from that of the exact many-body states.  

The non-unitarity of the overlap matrix is the reason why we do not use the boson representation in the calculation. In fact, we have developed a code to use the boson representation disregarding the non-linear terms of the transformation between the two boson bases and re-orthonormalizing the transformation. Consequently, the absolute values of some overlap matrix elements exceeded one by more than an order of magnitude.
Thus, this artificial unitarization method using the boson representation cannot be accepted.

\subsection{\label{subsec:identities} Identical initial and final states }
We assume that $|f\rangle=|i\rangle$ in this subsection. In this case, 
$\hat{v}^{(K^\prime \pi^\prime)\dagger}_I$  [Eq.~(\ref{eq:vhatI})] is equal to $\hat{v}^\dagger$ expressed as
\begin{eqnarray}
&&\hat{v}^\dagger = \sum_{m^{\prime\prime}} \hat{v}^{a\dagger}_{m^{\prime\prime}} \hat{v}^{b\dagger}_{m^{\prime\prime}}, \label{eq:vhatfei} \\
&&\hat{v}^{a\dagger}_{m^{\prime\prime}} = \sum_{\mu^\prime\nu^\prime}Y^{m^{\prime\prime}\ast}_{\mu^\prime\nu^\prime}
a_{\nu^\prime} a_{\mu^\prime}, \label{eq:vhat_a} \\
&&\hat{v}^{b\dagger}_{m^{\prime\prime}} = \sum_{\mu\nu}\left( \frac{1}{X} \right)^\ast_{\mu\nu,m^{\prime\prime}}
a_\nu a_\mu, 
\end{eqnarray}
where the suffixes $I$ and $F$ as well as $(K^\prime\pi^\prime)$ are omitted. The condition $K^\prime\neq 0$ is also assumed; however, 
this assumption is not essential. It can be shown in the QRPA order that
\begin{equation}
\left[ \hat{v}^{b\dagger}_{m^{\prime\prime}}, O^{\dagger}_{m^\prime} \right] = \delta_{m^{\prime\prime}m^\prime}. \label{eq:comm_va_O}
\end{equation}
Let us suppose that the backward amplitudes $Y_{\mu\nu}$ of the QRPA solutions $m$ and $m^\prime$ are very small.
Using Eq.~(\ref{eq:comm_va_O}) and ignoring 
$\langle i| O^\dagger_1\cdots O^\dagger_{2n}|i\rangle \sim Y^{n}$ with $n\geq 1$, we obtain 
\begin{eqnarray}
_I\langle m | m^\prime \rangle_I 
&\simeq& \frac{1}{{\cal N}^2} \langle i | O_m O^\dagger_{m^\prime} 
(1+\hat{v}^\dagger)(1+\hat{v}) | i \rangle \nonumber \\
&\simeq& \frac{1}{{\cal N}^2} \langle i | O_m O^\dagger_{m^\prime} | i \rangle 
\langle i | (1+\hat{v}^\dagger)(1+\hat{v}) | i \rangle \nonumber\\
&=& \delta_{mm^\prime}. \label{eq:ov_fei}
\end{eqnarray}
It is assumed in deriving the last expression that the normalization factor ${\cal N}^2$ is calculated up to the same order with respect to $\hat{v}$ 
as that of the denominator.
This derivation implies that the truncation of the $\hat{v}$-expansion does not affect the overlap matrix elements of 
the QRPA solutions that do not have the backward amplitudes. Thus, the non-collective states are expected to satisfy Eq.~(\ref{eq:ov_fei}) 
fairly  accurately. From our numerical calculations shown in Sec.~\ref{sec:result}, we confirmed that 
this expectation was correct for all of the non-collective states with a deviation less than $10^{-4}$. 
The deviation from Eq.~(\ref{eq:ov_fei}) with $m=m^\prime$ of the relatively  
collective real states among the QRPA solutions is around 0.01, and that of the two spurious states associated with the particle number is around 0.5. From this deviation, our method should be applied only to the cases for which the break in the particle number conservation is not so large that 
a large deviation from Eq.~(\ref{eq:ov_fei}) for the spurious states does not significantly affect the nuclear matrix elements that we finally require.


\section{\label{sec:calculation} Details of calculations}
\subsection{\label{subsec:numerical} Procedure}
We use the code of the HFB approximation explained in Refs.~\cite{Obe07, Bla05, Ter03} and that of the QRPA developed by us \cite{Ter10}. 
Although the two codes have been developed independently, in both codes, the wave functions are expressed in the cylindrical coordinates. The wave functions are interpolated by the B-splines, see e.g.~\cite{Nur89,Boo78}, and contained in a cylindrical box with the vanishing boundary condition. The HFB equation is solved with the cutoff at the quasiparticle energy of 20 MeV in order to avoid huge test calculations in terms of computational amount. The canonical quasiparticle basis \cite{Rin80} is used for the general quasiparticles in the formulation mentioned in  Sec.~\ref{sec:formulation}, and the basis wave functions are constructed so as to include the unbound components according to the method of Ref.~\cite{Ter05} after the HFB equation is solved. The unbound components are important for accurately obtaining the wave functions of the 
lesser occupied levels. 

Subsequently, trimming of the basis space is carried out by removing a small number of the 
canonical quasiparticle states with the least occupation probabilities in each $(q,\pi,j_z)$-subspace   
so that the dimension of the subspace is the same between the bases of the two nuclei if the corresponding original dimensions are different. 
This process is necessary in our calculation, because the dimension of each subspace is not a direct input to the HFB calculation, 
but the dimension is controlled by the cutoff quasiparticle energy. 

After this adjustment, the matrix elements of the unitary transformation are calculated according to Eqs.~(\ref{eq:TF1}) and (\ref{eq:TF2}). 
The two canonical quasiparticle spaces are not identical in the coordinate calculation if the corresponding dimensions are identical, 
because the truncated space is determined self-consistently by solving the HFB equation. Thus, a small correction is made to $T^{IF1}$ and $T^{IF2}$ in such a way that the canonical quasiparticle wave functions associated with the state $I$ obtained by the transformation (\ref{eq:transformation_qp}) are orthonormalized. The states with less occupation are mainly modified in this orthonormalization. 
The canonical quasiparticle wave functions are used for calculating $T^{IF1}$ and $T^{IF2}$ and the interaction matrix elements in the QRPA equation.

Subsequently, we calculate the matrix $D$ [Eq.~(\ref{eq:D})] and the normalization factor of the HFB state [Eq.~(\ref{eq:N_HFB})]. 
The $D$ matrix is singular, when the two HFB states are orthogonal as is seen in Eq.~(\ref{eq:N_HFB}), that is, the two nuclei share the same single-particle basis but have different configurations. This does not occur, however, as long as the HFB equations of the two nuclei are solved self-consistently. 


We solve the QRPA equation in the so-called matrix formulation \cite{Rin80,Ter05}.
The two-canonical-quasiparticle spaces used in the QRPA calculation are not truncated in the test calculations after the trimming. 
This treatment enables clear separation of the spurious states associated with the particle number. 
We reduce the size of the canonical-quasiparticle spaces using the tight cutoff of 20 MeV so that all allowed combinations of 
the two canonical quasiparticle states are easily tractable. Thus, the discussion of the nuclear properties is out of the scope of this paper. 
Using the QRPA solutions, one can calculate the matrix elements of the generator of the QRPA ground state [Eq.~(\ref{eq:CI_CF})] and the associated normalization factors [Eqs.~(\ref{eq:NI_QRPA}) and (\ref{eq:NF_QRPA})]. 
The matrix $X^{(K^\prime\pi^\prime)I}$ or $X^{(K^\prime\pi^\prime)F}$ would be singular, if all of the forward amplitudes of a QRPA solution vanish, or 
a two-canonical-quasiparticle component is not used in any of the QRPA solutions. However, this does not physically occur.

The next step is the calculation of the contractions (\ref{eq:contraction1})$-$(\ref{eq:contraction3}) using   
Eqs.~(\ref{eq:TF1}), (\ref{eq:TF2}), and (\ref{eq:D}). The generalized expectation values of high order with respect to the fermion operators are calculated by generating the list of the indices of the canonical quasiparticle states used in the contractions systematically and recursively from a low order (refer to the proof of Wick's theorem \cite{Wic50}). Finally, the overlap matrix elements (\ref{eq:fooi}) are calculated by using Eqs.~(\ref{eq:ft})$-$(\ref{eq:fot4}).

\subsection{\label{subsec:parallelization} Truncation scheme}
Feature of parallel computation affects the answer of a question of what approximation is efficient. 
Equation (\ref{eq:ft}) can be calculated by multiplication of three matrices having matrix elements $X^{Fm\ast}_{\mu\nu}$, $X^{Im^\prime}_{\mu^\prime\nu^\prime}$ and 
$\langle f | a^F_\nu a^F_\mu a^{I\dagger}_{\mu^\prime} a^{I\dagger}_{\nu^\prime}|i\rangle$. 
These matrices are partitioned, distributed to the cores of the computer, and handled by ScaLAPACK \cite{Sca12} in the process of the multiplication. 
However, this approach is not efficient in the calculation of  Eqs.~(\ref{eq:st})$-$(\ref{eq:fot4}). This is because the redistribution 
of the matrix elements of $C^{(K_1\pi_1)F\ast}$ and $C^{(K_1\pi_1)I}$ is necessary between the cores before the matrix 
multiplication is carried out using ScaLAPACK when different terms are calculated. 
This requires large computation times if the data size is large. 
Thus, we calculate the un-normalized overlap of 
Eqs.~(\ref{eq:st})$-$(\ref{eq:fot4}) truncating the two-canonical-quasiparticle states used in the summations without using ScaLAPACK (parallel computation is still used). 
The efficiency of this truncation is high due to the reason for the $\hat{v}$-expansion discussed in Sec.~\ref{sec:formulation}, 
that is,  $|I\rangle$ and $|F\rangle$ 
 have different configurations at the Fermi surface. This efficiency is confirmed numerically in Sec.~\ref{sec:result}. 
Obviously, this approximation also holds good for Eq.~(\ref{eq:ft}). 
However, that term is calculated without this approximation because the matrix multiplication using ScaLAPACK is very efficient.
We introduce another independent truncation of the two-canonical-quasiparticle states for the calculation of $F^1_{mm^\prime}$ 
(\ref{eq:fot1}) because those states $\{(\mu\nu)\}$ that are sufficient for the lower-order terms (\ref{eq:st}) and (\ref{eq:tt}) are too many to handle for the six-fold summation with respect to these states in the calculation of $F^1_{mm^\prime}$. 

Two truncations are used regarding $K\pi$; one is that in the summations of Eqs. (\ref{eq:st})$-$(\ref{eq:fot4}), and another is that of ${\cal M}$  [Eq.~(\ref{eq:M})]. These truncations are treated independently.
The last one is the truncation with respect to $\hat{v}^{(K_1\pi_1)\dagger}_F$ and $\hat{v}^{(K_1\pi_1)}_I$, as is shown in Eq.~(\ref{eq:fooi}). 
We calculate the normalization factor up to an order higher than that of the un-normalized overlap, as mentioned in Sec.~\ref{sec:formulation}. 
Thus, we use six truncations in the calculation of the overlap matrix after the canonical-quasiparticle bases are determined. These truncations are investigated numerically in Sec.~\ref{sec:result}.

\subsection{\label{subsec:properties}Properties of test states}
We discuss the physical properties of the test states that we use in this study. 
$^{26}$Mg and $^{26}$Si are used for $|i\rangle$ ($|I\rangle$) and $|f\rangle$ ($|F\rangle$), respectively, with the Skyrme parameter set SkM$^\ast$ \cite{Bar82} and the volume pairing density functional \cite{Dob96}. Two sets of test states 
are used with different pairing strengths. The properties of the HFB ground states are shown in Table~\ref{tab:26Mg_26Si}, and the pairing strengths are given in Table~\ref{tab:pairing_strength}. 
The total dimension of the HFB space is $\simeq$330 including those with negative $j_z$.
\begin{table}[h]
\caption{\label{tab:26Mg_26Si}%
Properties of HFB ground states of $^{26}$Mg and $^{26}$Si for test sets I and II. 
$\beta_\textrm{p}$ and $\Delta_\textrm{p}$ denote the quadrupole deformation and the averaged pairing gap of the 
protons. Those with the suffix n correspond to the quadrupole deformation and the averaged pairing gap of the neutrons.}
\begin{ruledtabular}
\begin{tabular}{ccccc}
 Nucleus & $\beta_\textrm{p}$ & $\Delta_\textrm{p}$ (MeV) & $\beta_\textrm{n}$ & $\Delta_\textrm{n}$ (MeV) \\
\colrule
\multicolumn{5}{c}{Test set I} \\
\colrule
$^{26}$Mg & $-0.199$ &0.794 &$-0.195$ & 1.510 \\
$^{26}$Si  & $-0.224$ & 0.865 & $-0.206$ & 1.402 \\
\colrule
\multicolumn{5}{c}{Test set II} \\
\colrule
$^{26}$Mg & $-0.228$ & 0.779 & $-0.234$ & $<$0.001 \\
$^{26}$Si  & \hspace{7.5pt}0.251 & 0.011 & \hspace{7.5pt}0.316 & \hspace{7pt}0.259 \\
\end{tabular}
\end{ruledtabular}
\end{table}
\begin{table}[h]
\caption{\label{tab:pairing_strength}%
Strengths of pairing energy functional $G_\textrm{p}$ and $G_\textrm{n}$ used.
A cutoff quasiparticle energy of 20 MeV is used in the HFB calculations.}
\begin{ruledtabular}
\begin{tabular}{ccc}
 Nucleus & $G_\textrm{p}$ (MeV$\;$fm$^3$) & $G_\textrm{n}$ (MeV$\;$fm$^3$) \\
\colrule
\multicolumn{3}{c}{Test set I} \\
\colrule
$^{26}$Mg & $-150.0$ & $-270.0$ \\
$^{26}$Si  & $-270.0$ & $-200.0$ \\
\colrule
\multicolumn{3}{c}{Test set II} \\
\colrule
$^{26}$Mg & $-150.0$ & $-150.0$ \\
$^{26}$Si  & $-150.0$ & $-150.0$ \\
\end{tabular}
\end{ruledtabular}
\end{table}

It is our intention to test two fairly different cases; in one case, the two nuclei have similar properties except for the difference in the proton and neutron numbers, and in another case, the two nuclei have fairly different properties. This difference can siginificantly affect the overlap matrix,  because the matrix is not unitary, i.e., there is no normalization of the matrix elements. 

It is a physical feature of the region around $^{26}$Mg that the sign of the ground-state quadrupole deformation is sensitive to the input parameters. We confirmed that the HFB ground states were axially symmetric using a three-dimensional HF-plus-BCS code as long as SkM$^\ast$ is used. 

\begin{figure}[h]
\includegraphics[width=8cm]{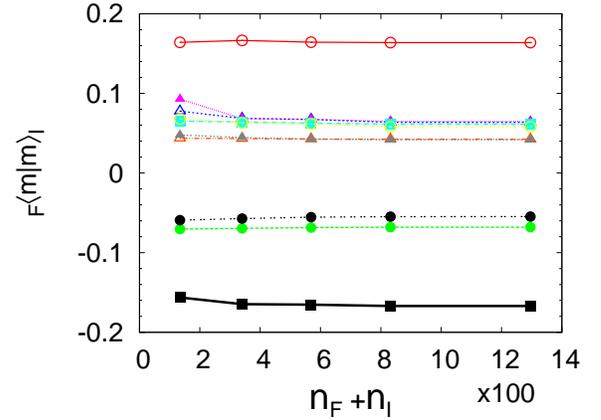}
\caption{ \label{fig:fNF+fNI} (Color online) Ten diagonal overlap matrix elements having largest absolute values as functions of $\mathfrak{N}_F+\mathfrak{N}_I$. }
\end{figure}

\begin{figure}[h]
\includegraphics[width=8cm]{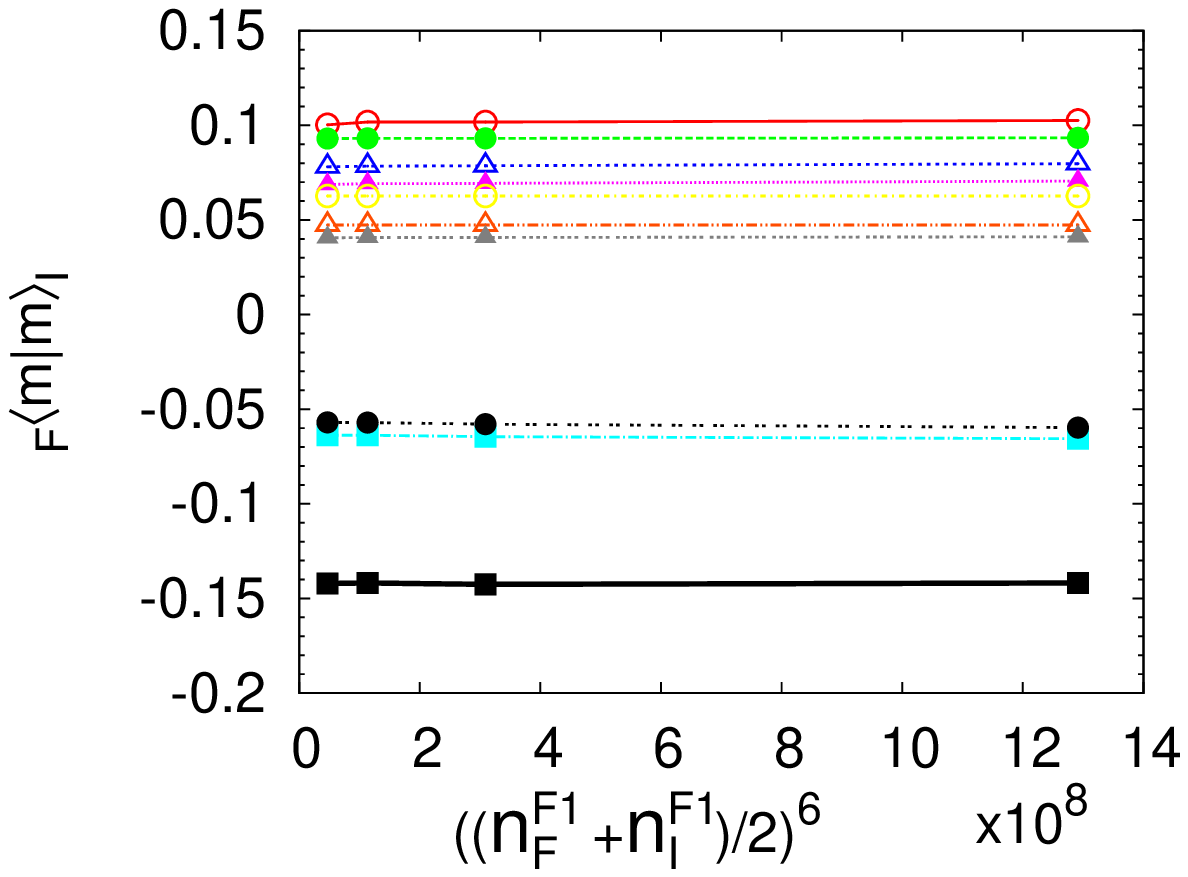}
\caption{ \label{fig:fNF1F+fNF1I} (Color online) Ten diagonal overlap matrix elements having largest absolute values as functions of $((\mathfrak{N}^{F1}_F+\mathfrak{N}^{F1}_I)/2)^6$. }
\end{figure}

\begin{figure}[h]
\includegraphics[width=8cm]{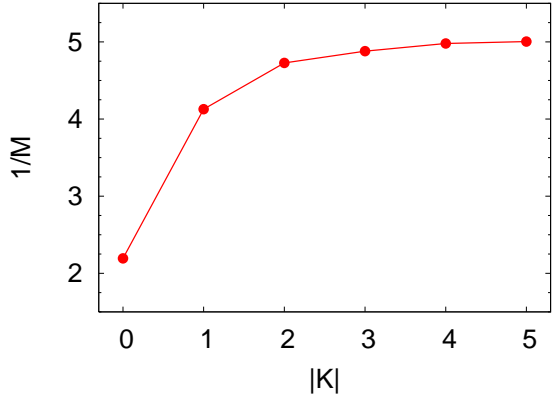}
\caption{ \label{fig:M} (Color online) Convergence of $1/{\cal M}$, Eq.~(\ref{eq:M}), with respect to $|K|$. Each value includes 
the contributions of both $\pi=+$ and $-$, and the terms with $K<0$ are also included.}
\end{figure}

\begin{figure}[h]
\includegraphics[width=8cm]{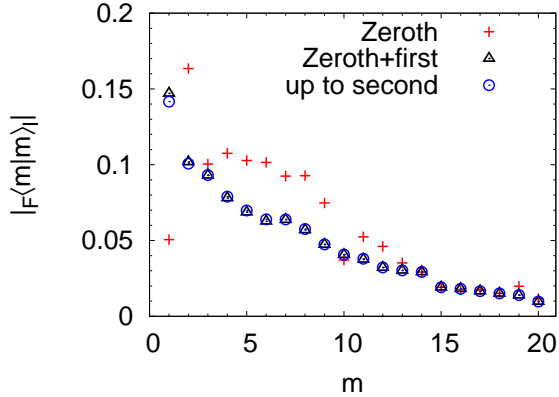}
\caption{ \label{fig:abs_ov_diag_I} (Color online) Twenty largest absolute values of diagonal overlap matrix elements. Those up to the second order with respect to $\hat{v}^{(0+)}_I$ and $\hat{v}^{(0+)}_F$ are shown in descending order. The terms with $(K_1\pi_1)\neq (0+)$ of Eq.~(\ref{eq:fooi}) are not included. 
We used $\mathfrak{N}_F+\mathfrak{N}_I=134$ and   
$((\mathfrak{N}^{F1}_F+\mathfrak{N}^{F1}_I)/2)^6=3\times 10^8$ (see Figs.~\ref{fig:fNF+fNI} and 
\ref{fig:fNF1F+fNF1I}).  
}
\end{figure}

\begin{figure}[h]
\includegraphics[width=8cm]{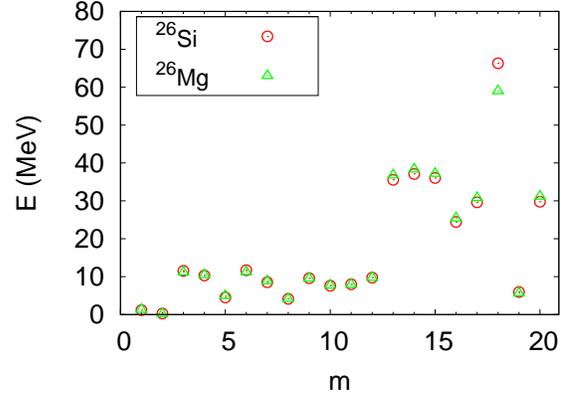}
\caption{ \label{fig:qrpa_e} (Color online)  Energies of $K^\pi=0^+$ QRPA excited states of $^{26}$Mg and $^{26}$Si 
in order corresponding to data in Fig.~\ref{fig:abs_ov_diag_I}. }
\end{figure}

\begin{figure}[h]
\includegraphics[width=8cm]{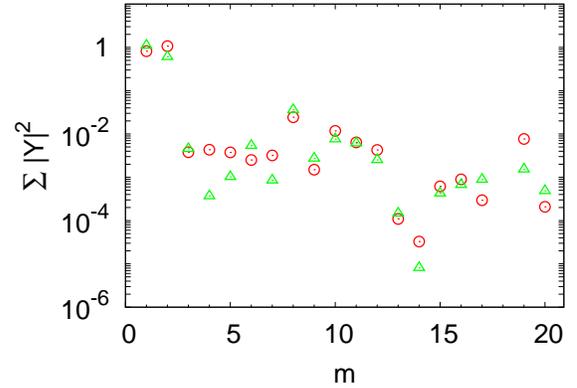}
\caption{ \label{fig:qrpa_sumy2} (Color online)  Summation of squared backward amplitudes of QRPA excited 
states corresponding to Fig.~\ref{fig:qrpa_e}.  The definition of the symbols is the same as that of Fig.~\ref{fig:qrpa_e}. }
\end{figure}

\begin{figure}[h]
\includegraphics[width=8cm]{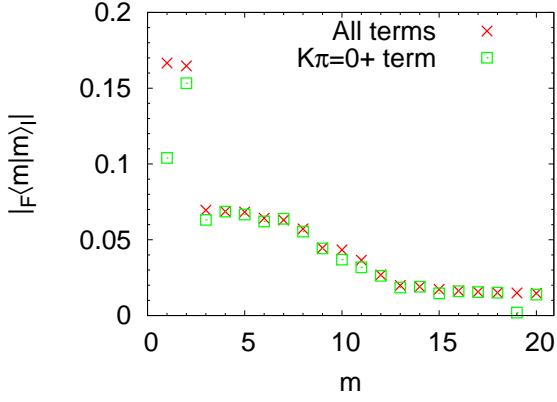}
\caption{ \label{fig:abs_ov_diag_kpnokp} (Color online)  Twenty largest absolute values of diagonal matrix elements of overlap in descending order with all values of $(|K_1|\pi_1)$, that is  (0+)$-$(4+), and those with only (0+) (not necessarily in descending order), see Eq.~(\ref{eq:fooi}). 
The converged ${\cal M}$ value is used for both calculations. Negative parity contributions are not included (see text). 
The QRPA state $m=1$ (2) corresponds to $m=2$ (1) of Fig.~\ref{fig:abs_ov_diag_I}. }
\end{figure}

\begin{figure}[h]
\includegraphics[width=8cm]{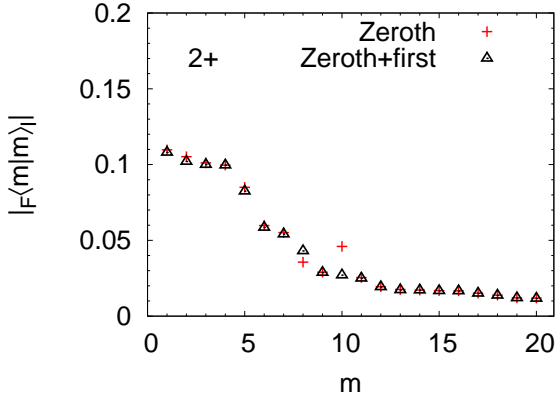}
\caption{ \label{fig:abs_ov_diag_2+} (Color online)  Twenty largest absolute values of diagonal matrix elements of overlap in descending order for  QRPA states of $(K\pi)=(2+)$ with only zeroth-order terms with respect to $\hat{v}^{(2+)}_I$ and $\hat{v}^{(2+)}_F$ and those also including  first-order terms. The terms with $(K_1\pi_1)\neq (2+)$ in Eq.~(\ref{eq:fooi}) are not included. Truncation was made at  $\mathfrak{N}_F+\mathfrak{N}_I=350$.
 }
\end{figure}

\begin{figure}[h]
\includegraphics[width=8cm]{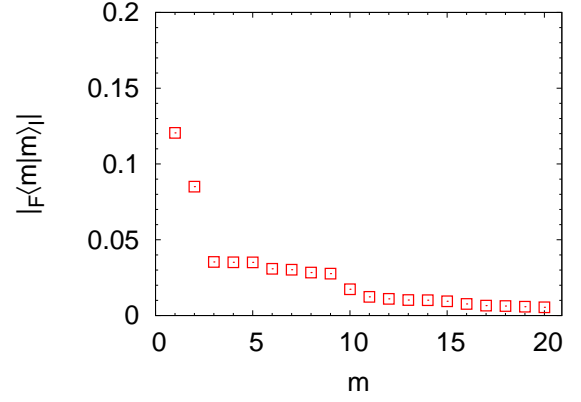}
\caption{ \label{fig:abs_ov_diag_spurious_test} (Color online)  The same as $K\pi=0+$ term in Fig.~\ref{fig:abs_ov_diag_kpnokp} but for the terms proportional to $_q\langle f|i \rangle_q$, $q$ denoting proton or neutron, 
not included.
 }
\end{figure}

\section{\label{sec:result} Numerical test of truncations}
\subsection{\label{subsec:test_state_set_I} Test set I }

We examine the effects of the various truncations separately using test set I with $(K\pi)=(0+)$. 
The convergence with respect to the number of the two-canonical-quasiparticle states used in Eqs.~(\ref{eq:st}) and (\ref{eq:tt}) is examined without the second-order term with respect to $\hat{v}^{(K\pi)}_I$ and $\hat{v}^{(K\pi)}_F$ [Eq.~(\ref{eq:fot})]. 
The truncation of the two-canonical-quasiparticle states for $F^{1}_{mm^\prime}$ [Eq.~(\ref{eq:fot1})] is investigated by suppressing $(K_1\pi_1)
\neq (0+)$ in the un-normalized overlap [$F^4_{mm^\prime}$, Eq.~(\ref{eq:fot4}), is omitted for simplicity].  
The effect of Eq.~(\ref{eq:fot}) is also investigated using  only $(K_1\pi_1)=(0+)$ in the un-normalized overlap.  
On the other hand, when the terms with $(K_1\pi_1)\neq (0+)$ are included, Eq.~(\ref{eq:fot}) is omitted.

Let $\mathfrak{N}_F$ and $\mathfrak{N}_I$ be the number of two-canonical-quasiparticle states associated with $|F\rangle$ and $|I\rangle$ truncated for calculating Eqs.~(\ref{eq:st}), (\ref{eq:tt}), and  (\ref{eq:fot2})$-$(\ref{eq:fot4}). Since $F^1_{mm^\prime}$, Eq.~(\ref{eq:fot1}), 
has six-fold summations with respect to $(\mu\nu)$, it is truncated separately as mentioned before.
We show the convergence of the overlap matrix elements with respect to $\mathfrak{N}_F+\mathfrak{N}_I$ in Fig.~\ref{fig:fNF+fNI}. 
$\mathfrak{N}_F$ and $\mathfrak{N}_I$ are controlled in the numerical calculation by using a cutoff occupation probability of the 
canonical quasiparticle states, and the states with larger occupation probabilities than the cutoff are used. 
The occupation probability is defined by the norm of the lower component of the quasiparticle wave function, and in our calculation, 
it is equal to the occupation probability of the canonical state. 
The same value of the cutoff is applied for the two bases, and we have  $\mathfrak{N}_F\simeq\mathfrak{N}_I$. It is seen that $\mathfrak{N}_F+\mathfrak{N}_I=350$ is sufficient for the convergence. The total number without the truncation is $\simeq$3300, and thus, 
this truncation is fairly efficient, as has been discussed before.

Figure \ref{fig:fNF1F+fNF1I} illustrates the convergence of the diagonal overlap matrix elements with respect to the number of the two-canonical-quasiparticle states used for the calculation of $F^{1}_{mm^\prime}$ (\ref{eq:fot1}). 
The term $\mathfrak{N}^{F1}_F$ is the number of those states associated with $|F\rangle$, and $\mathfrak{N}^{F1}_I$ is that  associated with $|I\rangle$. 
The terms $((\mathfrak{N}^{F1}_F+\mathfrak{N}^{F1}_I)/2)^6$ is the number of the terms of the six-fold summation 
with respect to $(\mu\nu)$ of  Eq.~(\ref{eq:fot1}) [only ($K_1\pi_1)=(0+)$]. A fairly stable convergence is obtained, and the value of 
$((\mathfrak{N}^{F1}_F+\mathfrak{N}^{F1}_I)/2)^6=5\times 10^7$ is sufficient for convergence. 
This implies that  $\mathfrak{N}^{F1}_F$ is at most 20 with $\mathfrak{N}^{F1}_F \simeq \mathfrak{N}^{F1}_I$, and thus, the second-order term (\ref{eq:fot}) can be considered as negligible.

The normalization factor $1/{\cal M} = {\cal N}_I{\cal N}_F$ does not have a mechanism of fast convergence with respect to   
$\mathfrak{N}_F+\mathfrak{N}_I$ unlike the un-normalized overlap matrix element, as mentioned before. 
Indeed, we found that no truncation was possible to satisfy Eq.~(\ref{eq:ov_fei}), thus, $1/{\cal M}$ was calculated without that truncation.
Figure \ref{fig:M} depicts the convergence of $1/{\cal M}$ with respect to $|K|$, thereby indicating that $|K|$ up to 3 is sufficient.
Further, we found that the normalization term up to the second order in Eq.~(\ref{eq:NI_QRPA}),
\begin{equation}
1+\sum_{K_1\pi_1}\langle i | \hat{v}^{(K_1\pi_1)\dagger}_I \hat{v}^{(K_1\pi_1)}_I | i \rangle,
\end{equation}
was 3.843, and that for $|f\rangle$ was 4.053. The fourth-order term for $|i\rangle$ [Eq.~(\ref{eq:fourth_norm_i})]
was 0.980, and that for $|f\rangle$ was 0.838. 
The first term of Eqs.~(\ref{eq:fourth_norm_i}) and (\ref{eq:fourth_norm}) (called as the unlinked term) was found to be larger than the second term by a factor 2$-$3. 
Thus, order estimation is possible for the normalization term of the sixth order 
\begin{equation}
\frac{1}{36}\sum_{K_1\pi_1}\langle i| \left( \hat{v}^{(K_1\pi_1)_I\dagger}_I\right)^3 \left( \hat{v}^{(K_1\pi_1)}_I\right)^3 |i \rangle,
\label{eq:6th_norm}
\end{equation}
which is not included in the calculations of the overlap, by considering the unlinked terms included in Eq.~(\ref{eq:6th_norm})
\begin{equation}
\frac{1}{36}\sum_{K_1\pi_1}\left( \textrm{Tr}(C^{(K_1\pi_1)}C^{(K_1\pi_1)\dagger})\right)^3 \times
\left\{
\begin{array}{rr}
6, & K\neq 0, \\
48, & K = 0.
\end{array} \right. \label{eq:sixth_norm}
\end{equation}
Equation (\ref{eq:sixth_norm}) gives a value of 0.088 for $|i\rangle$ and 0.113 for $|f\rangle$. We ignore this order of contributions.
The exchange terms in the second-order normalization terms (\ref{eq:exc_v2_nrm_i}) and (\ref{eq:exc_v2_nrm_f}) were found to be 
around $-$0.13 for $(K\pi)=(0+)$, and all of the absolute values of the terms with other $(K\pi)$ were smaller than 0.01 
with the tendency that the larger was the value of $|K|$, the smaller were the absolute values. These terms are also negligible.

The major diagonal overlap matrix elements are shown in Fig.~\ref{fig:abs_ov_diag_I}. It is seen that the contribution of the second-order term (\ref{eq:fot}) is negligible, and that of the first-order terms (\ref{eq:st}) and (\ref{eq:tt}) are not  significant for the small matrix elements. The zeroth-order term (\ref{eq:ft}) is sufficient in most of the matrix elements omitted in that figure.

Figure \ref{fig:qrpa_e} shows the energies of the QRPA excited states with $(K\pi)=(0+)$ of $^{26}$Mg and $^{26}$Si. It is seen from Figs.~\ref{fig:abs_ov_diag_I} and \ref{fig:qrpa_e} that the major diagonal matrix elements
of the overlap arise from the states with energy lower than 15 MeV. 
The charge symmetry of the two nuclei is also obvious\textemdash{}it is perfect in the low-energy region. 
Figure \ref{fig:qrpa_sumy2} shows the summations of the squared backward amplitudes of the QRPA solutions. 
These energies and backward amplitudes indicate that the states $m=1$ and 2 are the spurious solutions associated with the particle number. 

\begin{table}[h]
\caption{\label{tab:kpi_ne_0+}%
Contribution of $(|K_1|\pi_1)$ to first-order term [ Eq.~(\ref{eq:st}) plus Eq.~(\ref{eq:tt}) ] for 
$m=1$, see Fig.~\ref{fig:abs_ov_diag_kpnokp}. 
The summation of these contributions is equal to $0.32\times 10^{-2}$, and the summation except for $(K_1\pi_1)=(0+)$ 
is $6.24\times 10^{-2}$.
}
\begin{ruledtabular}
\begin{tabular}{cc}
 $|K_1|\pi_1$ & Contribution to the first-order term \\
 & ($\times 10^{-2}$)\\
\colrule
$0+$ & $-5.93$\hspace{5.0pt} \\
$1+$ &  \hspace{3pt}3.26 \\
$2+$ &  \hspace{3pt}0.47 \\
$3+$ &  \hspace{3pt}0.63 \\
$4+$ &  \hspace{3pt}1.90 \\
$5+$ &  $-0.002$ \\
\end{tabular}
\end{ruledtabular}
\end{table}

The contribution of $(K_1\pi_1) \neq (0+)$ to the first-order terms (\ref{eq:st}) and (\ref{eq:tt}) of the major overlap matrix elements are shown in Fig.~\ref{fig:abs_ov_diag_kpnokp}. We calculated that of $(K_1\pi_1)=(0-)$ and $(1-)$ and found that it was smaller than that of the positive parity by at least an order of magnitude, thus, only the positive parity is  used. The contribution of $(K_1\pi_1) \neq (0+)$ is very small to all the diagonal matrix elements except for those of the spurious states. 
This extreme sensitivity of the overlap of the spurious states provides us with one more reason why our 
method should be applied only to the cases for which the break in the particle number conservation is small.
Table \ref{tab:kpi_ne_0+} summarizes the details of the contribution  
to the most sensitive matrix element. The $(|K|\pi)$-dependence is irregular; however, eventually the contribution becomes negligible for the large value of $|K|$.
We also examined the contributions of the quasi-boson terms and the exchange terms of $\langle f|\hat{v}^{(K\pi)\dagger}_FO_m O^\dagger_m|f\rangle$. 
The absolute value of the quasi-boson term is larger than that of the exchange term by a factor 2$-$10 in many of the major matrix elements. 
Thus, the quasi-boson term is the leading term.

We also calculated the overlap matrix elements of the $(K\pi)=(2+)$ states (Fig.~\ref{fig:abs_ov_diag_2+}).
The result of the zeroth-plus-first order term can be compared with that of the $(K\pi)=(0+)$ term of Fig.~\ref{fig:abs_ov_diag_kpnokp}. The several largest values of $(K\pi)=(2+)$ are 50$-$60 \% larger than those of the real states of $(K\pi)=(0+)$, and the small ones in the tail 
of the curve are comparable.   
We discuss the origin of the overlap of the $(K\pi)\neq (0+)$ states based on the zeroth-order term 
(\ref{eq:ft}). The HFB wave function can be expressed as a direct product of the proton and neutron wave functions, that is, 
\begin{eqnarray}
&& | f \rangle = |f\rangle_p\otimes |f\rangle_n,\\
&& | i \rangle = |i\rangle_p\otimes |i\rangle_n, 
\end{eqnarray}
where $|f\rangle_p$ $(|i\rangle_p)$ and $|f\rangle_n$ $(|i\rangle_n)$ denote the proton and neutron wave functions. 
Since $\mu$ and $\nu$ ($\mu^\prime$ and $\nu^\prime$) in Eq.~(\ref{eq:ft}) are like particles, 
the generalized expectation value used in that equation is written as
\begin{widetext}
\begin{equation}
\langle f | a^F_\nu a^F_\mu a^{I\dagger}_{\mu^\prime}a^{I\dagger}_{\nu^\prime} | i \rangle
= \left\{
\begin{array}{ll}
_p\langle f | a^F_\nu a^F_\mu | i \rangle_p{}\ _n\langle f | a^{I\dagger}_{\mu^\prime} a^{I\dagger}_{\nu^\prime} 
| i \rangle_n , & \mu\nu\!\!: \textrm{protons and} \\
 & \mu^\prime \nu^\prime\!\!: \textrm{neutrons}, \\
_p\langle f | a^F_\nu a^F_\mu a^{I\dagger}_{\mu^\prime} a^{I\dagger}_{\nu^\prime} | i \rangle_p
{}\ _n\langle f | i \rangle_n, & \mu\nu\mu^\prime\nu^\prime\!\!: \textrm{protons}, \\
\textrm{the same equations but with} & \\ 
\textrm{the protons and neutrons exchanged.} & 
\end{array}
\right. \label{eq:gen_exp_pn}
\end{equation}
\end{widetext}
The $K$-quantum number of $|f\rangle_q$ and $|i\rangle_q$ ($q=p$ or $n$) is zero, 
thus, the first term of Eq.~(\ref{eq:gen_exp_pn}) vanishes for $K$ values of the QRPA state other than zero. 
Therefore, the overlap matrix elements of Fig.~\ref{fig:abs_ov_diag_2+} arise from the break 
in the particle-number conservation. 

Subsequently, we calculated the overlap matrix of the $(K\pi)=(0+)$ states without including the terms proportional to 
$_q\langle f | i \rangle_q$, and the corresponding results are shown in Fig.~\ref{fig:abs_ov_diag_spurious_test}. 
Upon comparing the largest overlaps except for those of the spurious states observed in Fig.~\ref{fig:abs_ov_diag_spurious_test} 
and those corresponding to the result labeled $(K\pi)=(0+)$ in Fig.~\ref{fig:abs_ov_diag_kpnokp}, 
it is observed that 55\% of the overlaps of the $(K\pi)=(0+)$ states arise from the equations 
that do not vanish in the case where the particle number is conserved. 
However, it is to be noted that removing the terms proportional to $_q\langle f|i\rangle_q$ artificially is not justified because the 
consistency between the equations derived is ignored. The optimal approach is to carry out the particle number projection of 
the many-body wave functions; however, this is out of the scope of this paper. 

\begin{figure}[h]
\includegraphics[width=8cm]{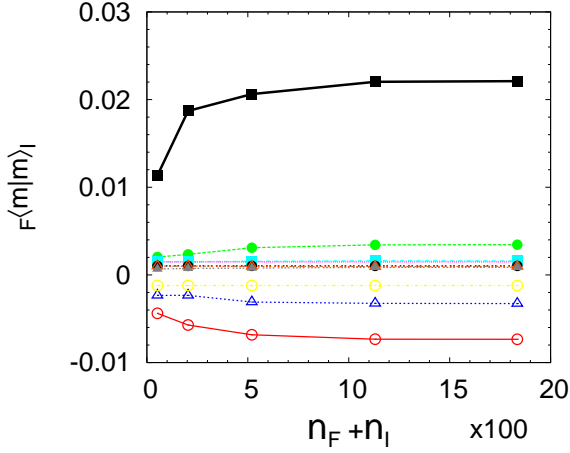}
\caption{ \label{fig:fNF+fNI_II} (Color online) Ten diagonal overlap matrix elements having largest absolute values 
as functions of $\mathfrak{N}_F+\mathfrak{N}_I$ obtained using test set II. The terms with $(K_1\pi_1)\neq (0+)$ are included in the 
first-order terms (\ref{eq:st}) and (\ref{eq:tt}), and the second-order term (\ref{eq:fot}) is not included.}
\end{figure}

\begin{figure}[h]
\includegraphics[width=8cm]{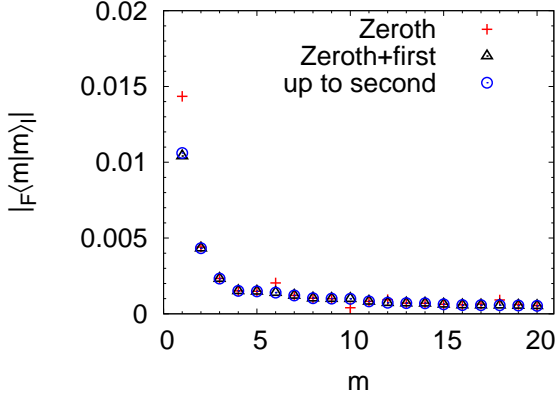}
\caption{ \label{fig:abs_ov_diag_II} (Color online) Twenty largest absolute values of diagonal overlap matrix elements obtained using test set II. The terms with $(K_1\pi_1)\neq (0+)$ are not included in the first- [ Eqs.~(\ref{eq:st}) and (\ref{eq:tt}) ] and second-term (\ref{eq:fot}), and the condition $\mathfrak{N}_F+\mathfrak{N}_I$ = 200 is used.}
\end{figure}

\begin{figure}[h]
\includegraphics[width=8cm]{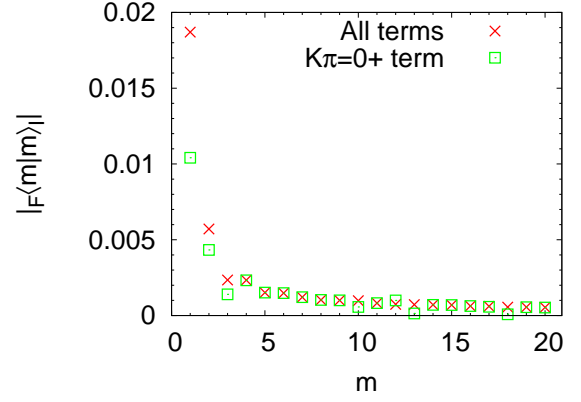}
\caption{ \label{fig:abs_ov_diag_kpnokp_II} (Color online) Twenty largest absolute values of diagonal overlap matrix elements obtained with and without $(K_1\pi_1)=(1+)$$-$$(5+)$ using test set II. The condition $\mathfrak{N}_F+\mathfrak{N}_I$ = 200 is used, and the second-order term (\ref{eq:fot}) is not included.}
\end{figure}

\begin{figure}[h]
\includegraphics[width=8cm]{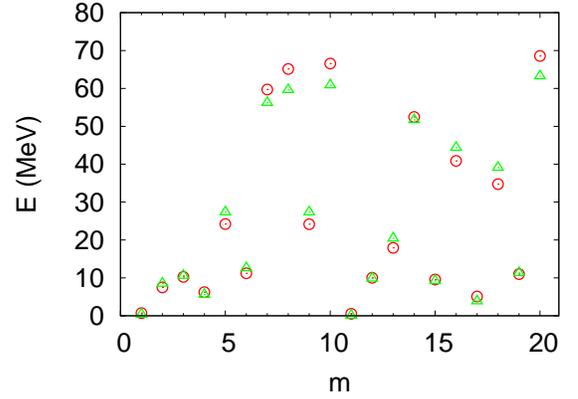}
\caption{ \label{fig:qrpa_e_II} (Color online) Energies of $K^\pi=0^+$ QRPA excited states of test set II. For the definition of the symbols, 
see Fig.~\ref{fig:qrpa_e}. }
\end{figure}

\begin{figure}[h]
\includegraphics[width=8cm]{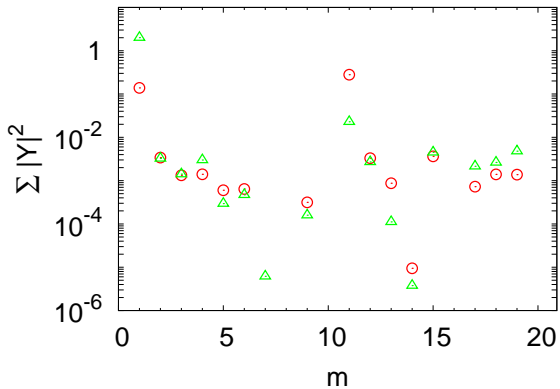}
\caption{ \label{fig:qrpa_sumy2_II} (Color online)  Summation of squared backward amplitudes of QRPA excited 
states corresponding to Fig.~\ref{fig:qrpa_e_II}.  For the definition of the symbols, see Fig.~\ref{fig:qrpa_e}. }
\end{figure}

\subsection{\label{subsec:test_state_set_II} Test set II}
In this subsection, we discuss the results obtained using test set II (Table~\ref{tab:26Mg_26Si}). Figure \ref{fig:fNF+fNI_II} illustrates 
the $\mathfrak{N}_F+\mathfrak{N}_I$-dependence of the major overlap matrix elements. The convergence is slow compared to that of test set I shown in Fig.~\ref{fig:fNF+fNI}. The possible reason is that the canonical quasiparticle basis is rather different between the two nuclei, as is seen in Table~\ref{tab:26Mg_26Si}, 
so that the basis wave functions with low occupation probability have certain components of the wave functions with larger occupation probabilities in another basis.

Figure \ref{fig:abs_ov_diag_II} shows the major diagonal matrix elements of the overlap. 
A couple of siginificant differences from Fig.~\ref{fig:abs_ov_diag_I} can be observed; one is that the values in Fig.~\ref{fig:abs_ov_diag_II} are  an order of magnitude smaller than those in Fig.~\ref{fig:abs_ov_diag_I}. 
This implies that if the two ground states are quite different, the overlap matrix elements become very small. The values of ${\cal S}$ defined by Eq.~(\ref{eq:S}) are $\sim$$10^{-2}$ (test set I) and $\sim$$10^{-5}$ (test set II) with $\textrm{dim}(G)\simeq 1650$. The second-order term (\ref{eq:fot}) and the terms with $(K_1\pi_1)\neq (0+)$ in the first-order terms (\ref{eq:st}) and (\ref{eq:tt}) are not included in these calculations. The value of ${\cal S}$ confirms that the all the matrix elements of the overlap are reduced significantly when the difference in the structure of the ground states of the two nuclei increases. 

It is noteworthy to compare the above values with that of the simple model discussed in Sec.~\ref{sec:analytical}. 
The ${\cal S}$ value of that model is 2/1650 $\sim$ $10^{-3}$, and this value is located between those of test sets I and II.
This order implies that the ground states of the two nuclei of test set I are closer to each other than those of the simple model, 
and those of test set II differ considerably from those of the simple model. 
The probable reason for the former relation is  the single-particle-configuration mixing arising from the similar pair fields of test 
set I (some components of the nuclear wave functions should be shared), 
and that for the latter relation is the quite different mean fields of test set II (the order of the single-particle levels is different).
A crude guideline is given for the overlap matrix from this argument as 
\begin{equation}
-\log (\textrm{dim}G) - 2 \alt
\log {\cal S} \alt
-\log (\textrm{dim}G) + 1.
\end{equation}

Another difference from the result of test set I is that the zeroth-order term (\ref{eq:ft}) is sufficient in many of the overlap matrix elements. The energies and the summation of the squared backward amplitudes of the QRPA solutions are shown in Figs.~\ref{fig:qrpa_e_II} and \ref{fig:qrpa_sumy2_II}, respectively. 
The results indicate that the QRPA solutions $m=1$ and 11 are the spurious states associated with the particle number. Upon comparing Figs.~\ref{fig:qrpa_sumy2} and \ref{fig:qrpa_sumy2_II}, we observe that the summation of the squared backward amplitudes of test set II is smaller on average than that of test set I except for the spurious states. Thus, the difference in the effect of $\hat{v}^{(K_1\pi_1)}_F$ and $\hat{v}^{(K_1\pi_1)}_I$ can be explained by the backward amplitudes [see Eq.~(\ref{eq:CI_CF})].

\section{\label{sec:summary} Summary}
The overlap matrix elements of the QRPA states based on the ground states of different nuclei have been calculated using the QRPA ground states explicitly. Our idea for handling the QRPA ground state is to expand it with respect to the generator---this approach appears to be a highly feasible method, and the expansion is probably the only feasible method. Further, certain analytical properties of the overlap matrix have been discussed. The non-unitarity is the exotic mathematical property of the overlap matrix in the QRPA. The truncation scheme used is explained in relation to the parallel computation, and the calculations are performed using relatively light nuclei with the Skyrme and the contact volume pairing energy functionals. 
The truncations of the calculation have been examined carefully and justified numerically. 

The computation provided the three following benefits: 
1) The truncation of the two-canonical-quasiparticle space is efficient in the calculation of the un-normalized overlap matrix. 
The normalization factor requires calculation with no truncation of the two-canonical quasiparticle space used in the QRPA calculation, 
and on the other hand, the calculation is reduced tremendously by using the identical bra and ket states, 
2) the inclusion of up to the linear term with respect to the generators of the QRPA ground state is sufficient in the expansion of the un-normalized matrix elements, 
and 3) the $(K_1\pi_1)\neq (K\pi)$ terms contribute negligibly to most of the un-normalized overlap matrix elements. 
The reason for the first benefit is obviously independent of nuclei, and the second benefit should be applicable to any nuclei 
for which the QRPA is a good approximation. 

As for the normalization factor, the maximum value of $|K| = 3$ is sufficient in our test calculations. 
The terms up to the fourth-order with respect to $\hat{v}^{(K_1\pi_1)}_I$ and $\hat{v}^{(K_1\pi_i)}_F$ were calculated in 
${\cal N}^2_I$ and ${\cal N}^2_F$, and the next-order terms were estimated to be negligible. 

Certain selection rules on the quantum numbers of the two-canonical quasiparticle states are used in the various terms, and  the terms with $(K_1\pi_1)\neq (K\pi)$ are subject to one more condition than the terms with $(K_1\pi_1) = (K\pi)$. 
Because of this difference, the former case has less terms than the latter case.
Thus, the terms of the former case seem less coherent, and this is the only explanation as regards the third benefit.  
Therefore, if the above explanation is correct, all of these benefits should also hold for heavier nuclei.  
Considering these advantages and the recent development of powerful parallel computers, there is no reason to avoid performing calculations  using the explicit QRPA ground state if accurate calculation is necessary.
Two sets of nuclear wave functions were used, and it has been shown that the overlap was sensitive to the difference in the wave functions of the
initial and final states. The feasibility of the calculation has been demonstrated in both the test cases.

We have included as many terms as possible in our calculations. As a result of this manner, certain non-quasi-boson (exchange) terms are included, while others are not.
The generators of the QRPA ground state were obtained by the quasi-boson approximation. 
It does not seem possible to extract the matrix element $C^{(K_m\pi_m)I}_{\mu\nu,\mu^\prime\nu^\prime}$ in the isolated form unless the exchange terms are ignored.
According to our experience of the calculation of the equations including both the quasi-boson and exchange terms, 
the exchange terms are not as significant as the quasi-boson terms perhaps because of a reason similar to that for the third benefit discussed above. 

The code is developed in such a manner that memory-shortage issues do not occur if applied to heavy nuclei, 
and the parallelization efficiency is good. Thus, the applicability of our method to heavy nuclei is a matter of availability of core-hours.
We are preparing to apply our method to the calculation of the nuclear matrix elements including the phase-space factor 
for a dozen of the $0\nu\beta\beta$ decays. 
Finally, it should be possible to apply our method to calculate the overlap to the pn-QRPA.

\begin{acknowledgments}
We are grateful to Dr.~Engel for suggesting the application of the like-particle QRPA to the nuclear matrix elements of the $0\nu\beta\beta$ decay and for the useful discussions.
We thank Drs. Oberacker and Umar for letting us use their HFB code.
This research is supported by the Ministry of Education, Culture, Sports, Science and Technology, Japan under the Grant-in-Aid for Scientific Research No.~23840005 
and HPCI Strategic Program Field 5. 
The computers used for our calculations are those at the Center for Computational Sciences, University of Tsukuba, under the Collaborative Interdisciplinary Program and the Computational Fundamental Science Project (T2K-Tsukuba); the Yukawa Institute for Theoretical Physics, Kyoto University (SR16000); and the Research Center for Nuclear Physics, Osaka University (SX-8R). 
We also used the computers at the National Institute for Computational
Sciences and the National Energy Research Scientific Computing Center during the early stage of this research.
\end{acknowledgments}

\end{document}